\documentclass[12pt]{article}
\usepackage{amsfonts,amssymb,amsmath}
 \usepackage{epic,eepic,cite}
\usepackage{theorem}
\newtheorem{definition}{Definition}[section]

\newtheorem{conjecture}[definition]{Conjecture}

\newtheorem{theorem}[definition]{Theorem}

\newtheorem{lemma}[definition]{Lemma}

\theorembodyfont{\rmfamily}
\newtheorem{rmk}{Remark}[section]

\numberwithin{equation}{section}

\newcommand{\nonu}{\nonumber \\}

\newcommand{\hs}[1]{\hspace{#1 mm}}

\newcommand{\eps}{\epsilon}

\newcommand{\vph}{\varphi}

\def\cA{{\cal A}}                    
\def\cD{{\cal D}}          \def\cE{{\cal E}}          
                    
\def\cJ{{\cal J}}                    \def\cL{{\cal L}}
          \def\cN{{\cal N}}

\newcommand{\bcN}{\overline \cN}

\def\fc{{\mathfrak c}}

\def\fs{{\mathfrak s}}
\def\ft{{\mathfrak t}}

\newcommand{\CC}{{\mathbb C}}

\newcommand{\II}{{\mathbb I}}

\newcommand{\RR}{\mbox{${\mathbb R}$}}

\newcommand{\ZZ}{{\mathbb Z}}

\newcommand{\wh}[1]{\widehat{#1}}
\newcommand{\wt}[1]{\widetilde{#1}}
\newcommand{\mb}[1]{\hs{4}\mbox{#1}\hs{4}}
\newcommand{\qmbox}[1]{{\qquad\mbox{#1}\quad}}
\newcommand{\half}{\frac{1}{2}}

\newcommand{\prf}{\underline{Proof:}\ }
\newcommand{\finprf}{\null \hfill {\rule{5pt}{5pt}}\\[2.1ex]\indent}

\newcommand{\atopn}[2]{\genfrac{}{}{0pt}{}{#1}{#2}}

\def\sump{\mathop{\sideset{}{'}\sum}}

\begin{document}
\renewcommand{\thefootnote}{\arabic{footnote}}
\setcounter{footnote}{0}
\newpage
\setcounter{page}{0}

\pagestyle{empty}

\null
\vfill
\begin{center}

{\Large \textsf{Thermodynamical limit of general $gl(N)$ spin chains II: 
\\[1.2ex]
Excited states and energies}}

\vspace{10mm}

{\large N. Cramp{\'e}$^{ab}$ , L. Frappat$^c$ and
{\'E}. Ragoucy$^c$}
\footnote{crampe@sissa.it,
luc.frappat@lapp.in2p3.fr, eric.ragoucy@lapp.in2p3.fr}

\vspace{10mm}
\emph{$^a$ International School for Advanced Studies}

\emph{Via Beirut 2-4, 34014 Trieste, Italy}
\vspace{5mm}\\

\emph{$^b$ Istuto Nazionale di Fisica Nucleare}

\emph{Sezione di Trieste}
\vspace{5mm}\\

\emph{$^c$ Laboratoire d'Annecy-le-Vieux de Physique Th{\'e}orique}

\emph{LAPTH, CNRS, UMR 5108, Universit{\'e} de Savoie}

\emph{B.P. 110, F-74941 Annecy-le-Vieux Cedex, France}

\end{center}

\vfill
\vfill

\begin{abstract}
We consider the thermodynamical limit of a $gl(N)$ spin chain with arbitrary 
representation at each site of the chain. We consider  
excitations (with holes and new strings) above the vacuum and 
compute their corrections 
in $\frac{1}{L}$ to the densities and the energy. 
\end{abstract}

\vfill
\vfill

\rightline{\texttt{arXiv:0710.5904}}
\rightline{LAPTH-1212/07}
\rightline{October 2007}

\baselineskip=16pt

\newpage

\pagestyle{plain}

\section{Introduction}
Integrable quantum spin chains are quantum mechanics models which 
can be solved exactly. Certainly, among them, 
the most studied is the spin $\half$
Heisenberg spin chain (also called the XXX chain) \cite{heisen} 
solved by H.~Bethe \cite{bethe} using his celebrated ansatz.
However, in recent applications as 
condensed matter experiments \cite{comp1,condmat,carbon} or 
string theory (for recent reviews, see \cite{BS,Agar} and references 
therein), more involved models show up.
 These models deal with higher rank Lie algebras 
such as $gl(N)$ (the XXX chain being associated to $gl(2)$) and/or
 higher spin (i.e. higher dimensional representation) on each site, 
see e.g.
\cite{ZAFA,Kul,Tak,Bab,ow,MENERI,KoIzBo,fafa,KuSu,Bytsko,Tsuboi}. 
They also include 
spin chains with impurities \cite{anjo,fuka,YWang} and
 alternating spin chains \cite{dewo,martins,abad2,ana}.
In most papers, each model is investigated on its own. Obviously, in 
opposition to these case by case studies, 
it would be of very much interest to develop a general framework 
which allows one to
solve all these models simultaneously. It is provided by the study
 of an `algebraic version' of spin chains which encompasses a generic 
 spin chain where each sites may carry different 
representations of $gl(N)$. The resolution of this general model
started with the computation of the Bethe equations in
\cite{KulResh,ow,byebye}. Let us note that a
very similar approach has been developped in \cite{MTV,TV} and 
\cite{paku} where an `algebraic version' of Bethe vectors is 
computed, in term of $L$-operator (first case) or Drinfeld currents 
(second case). Then, the vacuum state of this model was 
constructed in
\cite{thermy} in the thermodynamical limit (i.e. when the length $L$ of 
the chain tends to infinity) when the representations on the chain
are 
characterized by rectangular Young tableaux (of arbitrary height and 
width). The vacuum was proved to be of spin 0 and
formed by filled Fermi seas. 

In this paper, we carry on the study started in \cite{thermy} with 
the calculation of the first excited states 
above the vacuum state. This construction generalizes the results 
obtained for the fundamental $gl(N)$ spin chain \cite{sutherland,KulResh,anne}.
The plan of the paper is the following. We first remind in section 
\ref{sec:sum} the results obtained in \cite{thermy}. Let us stress 
that the rest of the 
paper heavily relies on the results given there. In section 
\ref{sec:holExcit}, we consider hole excitations above the 
vacuum state, and the $\frac{1}{L}$ corrections these excitations 
induce on the densities in the thermodynamical limit. 
We also compute the form of some hole   
excitations valid for any type of spin chains, as well as hole
configurations corresponding to `small' excitations, and valid for a 
large class of spin chains (including all $sl(2)$ spin chains and 
$gl(N)$ alternating ones). More general excitations where new strings 
are added to the vacuum configuration are dealt in section 
\ref{sec:gh}. The $\frac{1}{L}$ corrections to the densities are 
computed (theorem \ref{th:cj}) and we give a form for the 
excitation leading to a 
state with `small' (but non-trivial) spin. The Bethe equations relating 
the Bethe roots for holes and the ones for new strings are also 
computed (theorem \ref{th:constraint}). In section \ref{sec:energies}, we give a general form for a 
$L_{0}$-local Hamiltonian and compute the $\frac{1}{L}$ corrections 
to the energies for the excitations presented in sections \ref{sec:holExcit} 
and \ref{sec:gh}. Finally, appendix \ref{ap:proof} is devoted to the 
proof of theorem \ref{th:constraint}.

\newpage

\section{Notations and summary of previous results \label{sec:sum}}

We give, in this section, a review of the previous results such that 
this article be self-contained.

\subsubsection*{Bethe ansatz equations and string 
hypothesis\label{sec:notation}}

In the papers \cite{KulResh,ow,byebye}, the Bethe ansatz equations have been 
established for spin chains where each site may carry a different 
representation of $gl(N)$. To solve these equations, it is usual to use 
the string hypothesis which states that the solutions 
gather into $\nu_m^{(j)}$ strings of length $2m+1$ ($m\in\half\ZZ_+$)
of the following form
\begin{equation}
\lambda^{(j)}_{m,k}+i\,\alpha
\,,\quad \alpha=-m,-m+1,\dots,m \qmbox{and} j=1,\dots, N-1
\end{equation}
where $k=1,\dots,\nu^{(j)}_{m}$ and $\lambda^{(j)}_{m,k}$, the center of
the string, is real. 
Moreover, we will follow the lines given in \cite{thermy} and consider 
representations which have a rectangular 
Young tableau. They are thus
characterized by two integers $(a_{\ell},j_{\ell})$ for the site 
$\ell=1,\ldots,L$, where $a_{\ell}$ is the width of the Young tableau, 
and $j_{\ell}$ its height. The corresponding $sl(N)$ spin is 
$S=(\underbrace{0,\ldots,0}_{j_{\ell}-1},a_{\ell},
\underbrace{0,\ldots,0}_{N-1-j_{\ell}})$.

Under this restriction, the BAE for the centers of strings can 
be rewritten as
\begin{eqnarray}
\label{eq:bethe-argumt} &&\sum_{\ell=1}^L
\delta_{j,j_{\ell}}\,\Phi_{a_{\ell}}^{(m)}\left(\lambda_{m,k}^{(j)}\right) -
2\pi\,Q_{m,k}^{(j)}=
\label{eq:phasBAE}\\
&& \sum_{p\in\half\ZZ_+}\left(
\sum_{\ell=1}^{\nu^{(j-1)}_{p}}
\Phi_{-1}^{(p,m)}(\lambda^{(j)}_{m,k}-\lambda^{(j-1)}_{p,\ell})+
\sum_{\ell=1}^{\nu^{(j)}_{p}}
\Phi_{2}^{(p,m)}(\lambda^{(j)}_{m,k}-\lambda^{(j)}_{p,\ell})+
\sum_{\ell=1}^{\nu^{(j+1)}_{p}}
\Phi_{-1}^{(p,m)}(\lambda^{(j)}_{m,k}-\lambda^{(j+1)}_{p,\ell})
\right)\nonumber
\end{eqnarray}
where the parameters $Q_{m,k}^{(j)}$ are quantum numbers. 
We have introduced the functions:
\begin{eqnarray}
 \Phi_{p}^{(m)}(\lambda)&=& \sum_{\alpha=-m}^m \vph_{p+2\alpha}(\lambda)
\,,\ p\in\ZZ_{+}\ ,\ m\in\half\ZZ_{+}\qquad\\
 \Phi_{2}^{(m,n)}(\lambda) &=& 
\displaystyle \vph_{2m+2n+2}(\lambda)\,+
\,\vph_{2\vert m-n\vert}(\lambda)\,+2\sum_{\alpha=\vert m-n\vert+1}^{m+n}
\vph_{2\alpha}(\lambda)
\qquad\qquad m,n\in\half\ZZ_{+}\quad\quad\\
\Phi_{-1}^{(m,n)}(\lambda) &=& -\sum_{\alpha=\vert m-n\vert}^{m+n}
\vph_{2\alpha+1}(\lambda) \\
\vph_{p}(\lambda) &=& 2\,\mbox{Arctan}\left(\frac{2\lambda}{p}\right)\,,\
p\in\ZZ\,,\ p\neq0\mb{and}
\vph_{0}(\lambda) = 0\,.
\end{eqnarray}
A state is completely characterized by the data of the numbers of strings, $\nu^{(j)}_n$,
and the quantum numbers, $Q_{m,k}^{(j)}$. We can show that these quantum numbers are constrained 
by \cite{thermy}:
\begin{eqnarray}
&&Q_{m,k}^{(j)}\in[-Q_{m,max}^{(j)}\ ,\ Q_{m,max}^{(j)}]\\
&&\hspace{-0.8cm}Q_{m,max}^{(j)}=\half\Big(\nu^{(j)}_{m}-1
  +\sum_{\ell=1}^{L}\delta_{j,j_{\ell}}\,\min(2m+1, a_{\ell})
 -\sum_{n\in\half\ZZ_{+}}
\min(2m+1,2n+1)\,(2 \nu^{(j)}_{n}-\nu^{(j-1)}_{n}-\nu^{(j+1)}_{n})\Big)
\nonumber
\end{eqnarray}
The valence, $P_m^{(j)}$, is  the number of allowed quantum numbers i.e. 
$P_m^{(j)}=2Q_{m,max}^{(j)}+1$.

\subsubsection*{Regular spin chains: definition and notations}

We want to consider the thermodynamical limit (i.e. $L\rightarrow+\infty$) but
it seems impossible to deal with an infinite number of different representations.
Thus, following the computations done in \cite{thermy}, we will work with 
$L_{0}$-regular closed spin chains, that is to say periodic spin chain
(of length $L$) such that the representation (the $gl(N)$ spin) at 
site $\ell$ is the same as the one at site $\ell+L_{0}$, keeping 
$L_{0}$ finite when $L\rightarrow+\infty$.
Of course, for a $L_{0}$-regular spin chain, there is 
\textit{at most} $L_{0}$ different Young tableaux. 
However, among a sub-chain 
of length $L_{0}$, the same 
representation may appear several times. In this sub-chain, inequivalent representations may 
differ by 
distinct values of $j_{\ell}$ and/or $a_{\ell}$. 
We define $\cL\leq L_{0}$, the 
number of representations which have different values of $a_{\ell}$, 
and called $\bar a_{\alpha}$, $\alpha=1,\ldots,\cL$ these values. 
Accordingly, we define the ordered set: 
\begin{equation}
\cN=\left\{n_{\alpha}=\frac{\bar a_{\alpha}-1}{2}\mb{s.t.}
1\leq\alpha\leq\cL\right\}\subset\half\ZZ_{\geq0}
\mb{with} n_1<n_2<...< n_\cL\;.
\end{equation} 
As a convention, we will also introduce $n_{0}=-\half$ and $n_{\cL+1}=+\infty$.
The complementary set of $\cN$ is
\begin{equation}
\bcN=\half\ZZ_{\geq0}\setminus\cN=\left\{m\in\half\ZZ_{\geq0}\mb{s.t.} 
m\not\in\cN\right\}
\label{def:barN}
\end{equation} 
We also introduce the sets of indices defined by:
\begin{equation}
    \label{eq:set}
I_{\alpha}=\{\ell\in [1,L_{0}] \mb{s.t.} a_{\ell}=\bar a_{\alpha}\}\ ,\ \forall
\alpha\in[1,\cL]
\end{equation}
such that
\begin{equation}
\label{eq:s2s}
\sum_{\ell=1}^{L_{0}}\,(\ldots)_{\ell}\ =\
\sum_{\alpha=1}^{\cL}\,\sum_{\ell'\in I_{\alpha}}\,(\ldots)_{\ell'} \;.
\end{equation}
The cardinal $|I_{\alpha}|$ corresponds to the multiplicity of $\bar 
a_{\alpha}$ within a subset of $L_{0}$ sites. We define also
\begin{equation}
\cJ_{\alpha,j}=\sum_{\ell\in I_{\alpha}}\delta_{j,j_\ell}
\label{def:cJ-aj}
\end{equation}
which corresponds to the multipicity of the representation 
$(\bar a_\alpha,j)$ within a subset of $L_0$ sites. We thus have 
the property 
\begin{equation}
L_{0}=\sum_{\alpha=1}^{\cL} \sum_{j=1}^{N-1} \cJ_{\alpha,j}\,.
\label{eq:L0cJ}
\end{equation}
We will also call
$\cJ$ the greatest common divisor (gcd) of the 
$\cJ_{\alpha,j}$'s. In most of the cases (i.e. as soon as a 
representation appears only once in the subchain of length $L_{0}$), 
$\cJ$ is in fact equal to 1.

\subsubsection*{Bethe equations for the vacuum state}

Now, we can write down the Bethe equations in the thermodynamical limit for 
different states and, in particular, the vacuum state which is, by definition,
the state such that $P_m^{(j)}-\nu_m^{(j)}$ vanish for any $m$ and $j$. 
In \cite{thermy}, we prove that this defines the 
state uniquely and one can show that it has spin zero (i.e. it is a 
trivial $gl(N)$ representation).
This corresponds to the following choice of parameters
\begin{eqnarray}
\nu_{n_\alpha}^{(j)}\!\! &=&\!\!\frac{L}{NL_0} \sum_{\ell\in I_\alpha} \min(j,j_{\ell})\,
\Big(N-\max(j,j_{\ell})\Big)\qmbox{for $1\leq \alpha\leq\cL$ and}\nu_n^{(j)}=0
\qmbox{for $n\in\overline\cN$}\qquad
\label{string-vac}\\
Q_{m,k}^{(j)}\!\!&=&\!\!k-\half(\nu_m^{(j)}+1)\qmbox{for $k=1,\dots,\nu_m^{(j)}$ and $n\in\cN$}\;.
\end{eqnarray}

We will 
consider the thermodynamical limit $L\to\infty$, keeping $L_{0}$ 
finite, of the Bethe equations (\ref{eq:bethe-argumt}) 
for a $L_0$-regular spin chain and for the vacuum state. Then, we obtain the following 
set of equations for the densities, $\sigma_m^{(j)}(\lambda)$ ($m\in\cN$) of the center of
 the strings $\{\lambda_{m,k}^{(j)} \ |\ k=1,\dots,\infty \}$:
\begin{eqnarray}
&&\sum_{n\in\cN}\left\{
\int_{-\infty}^{\infty} d\lambda \,\sigma_{n}^{(j-1)}(\lambda)\,
{\Psi_{-1}^{(m,n)}}(\lambda_{0}-\lambda)
+\int_{-\infty}^{\infty} d\lambda \,\sigma_{n}^{(j)}(\lambda)\, 
{\Psi_{2}^{(m,n)}}(\lambda_{0}-\lambda)
\right.\nonu
&&\qquad
\left.+\int_{-\infty}^{\infty} d\lambda \,\sigma_{n}^{(j+1)}(\lambda)\,
{\Psi_{-1}^{(m,n)}}(\lambda_{0}-\lambda)
\right\}\  =\
-2\pi\,\sigma_m^{(j)}(\lambda_{0})+
\frac{1}{L_{0}}\sum_{\alpha=1}^{\cL}
\Big(\sum_{\ell\in I_\alpha}\delta_{j,j_{\ell}}\Big)\,
\Psi_{\bar a_{\alpha}}^{(m)}(\lambda_{0})\qquad\qquad\\
&&\qquad \forall \lambda_{0}\in\;
 ]-\infty\,,\,\infty[\ ,\ \forall
j=1,\ldots,N-1\ ,\ \forall m\in\cN\nonumber
\end{eqnarray}
where $\Psi(\lambda)$ are derivative of $\Phi(\lambda)$.

\subsubsection*{Densities for the vacuum state}

To solve this set of equations, we perform a Fourier transform, with the following 
choice for the normalisation 
\begin{equation}
\wh{f}(p)  = \frac{1}{2\pi}\int_{-\infty}^{\infty}e^{ip\lambda} \;
f(\lambda)\; d\lambda 
\end{equation}
and encompass them in a matrix. Finally, we obtain the following form of the BAE:
\begin{equation}
2\pi\,\wh\Psi(p)\,\wh\Sigma(p)=\Lambda(p)
\label{BAE-TF}
\end{equation}
where the $(N-1)\cL\times(N-1)\cL$ matrix 
$\wh\Psi(p)=-\cA(p) \otimes \wh\Psi_{-1}(p)$ 
and $\cA(p)$ is a $(N-1)\times(N-1)$ tridiagonal matrix with the non-vanishing entries
\begin{equation}
[\cA(p)]_{jj}=2\cosh\left(\frac{|p|}{2}\right)\qmbox{and}[\cA(p)]_{jj+1}=-1=\cA(p)_{j+1j}
\end{equation}
and $\wh\Psi_{-1}(p)$ is a $\cL\times\cL$ matrix such that
$[\wh\Psi_{-1}(p)]_{\alpha,\beta}=\wh\Psi^{(n_{\alpha},n_{\beta})}_{-1}(p)$.
We have also introduced the $(N-1)\cL$ vectors 
\begin{equation}\label{eq:SigLam}
\wh\Sigma(p)=\sum_{j=1}^{N-1}\sum_{\alpha=1}^{\cL}
\wh\sigma_{n_\alpha}^{(j)}\ e_j^{(N-1)}\otimes e_\alpha^{(\cL)}\qmbox{and}
\Lambda(p)=\frac{1}{L_0}\sum_{j=1}^{N-1}\sum_{\alpha,\beta=1}^{\cL}
\sum_{\ell\in I_\beta}\ \delta_{j,j_\ell}\wh\Psi_{\bar a_\beta}^{(n_\alpha)}(p)
\  e_j^{(N-1)}\otimes e_\alpha^{(\cL)}
\end{equation}
where $e_i^{(p)}$ is the canonical basis of $\CC^p$.
Then, inverting (\ref{BAE-TF}) and performing an inverse Fourier transform,
we get the densities of the centers of strings for the vacuum state:
\begin{eqnarray}
\sigma^{(k)}_{n_\alpha}(\lambda) &=& \frac{1}{N L_{0}}\sum_{\ell\in I_{\alpha}} 
 \ \ \sum_{q=(\vert k-j_\ell\vert+1)/2}^{(k+j_\ell-1)/2} \quad
\frac{\displaystyle \sin\Big(\frac{2q\pi}{N}\Big)}
{\displaystyle \cosh\Big(\frac{2\pi}{N}\lambda\Big) - 
\cos\Big(\frac{2q\pi}{N}\Big)}\qquad 1\leq\alpha\leq\cL
\label{eq:densitemagique}
\end{eqnarray}
These densities allow us to compute physical quanties such as the 
vacuum energy.

We will deal with sums that runs on integers or half-integers, 
 with an increment which can be either 1 or $\half$. To avoid 
 confusion between these two types of sums, we will denote with a 
 prime, $\sump$, the sums that increment by $\half$ step, keeping the 
 usual sum, $\sum$ for the ones with step 1. Hence, we have for 
 instance the relations (for $m$ integer):
\begin{eqnarray}
 \sump_{n=0}^{m} = \sum_{n=0}^{m} + \sum_{n=\half}^{m-\half}
 \mb{and}
 \sump_{n=\half}^{m+\half} = \sum_{n=1}^{m} + \sum_{n=\half}^{m+\half}
\end{eqnarray}

\null

The aim of the present paper is to consider the first excitations 
above the vacuum state (as defined above). 
There will be of two different types: 
\textit{holes}, which correspond to a configuration where some of the 
strings defining the vacuum state have been removed, and 
\textit{new strings} which may be added once holes have been done.
We will treat the BAEs for these excitations, that is to say gives 
(in the thermodynamical limit) the first corrections (in $\frac{1}{L}$) to the 
densities. We will also compute corrections to the energies and 
give `local' Hamiltonians.

\section{Excited states: holes \label{sec:holExcit}}

In this section, we consider the first excitations above the vacuum state. 
They are constructed from the vacuum by modifying the root distribution.
For such a purpose, we will create holes (i.e. remove some strings) in
the filled seas of the vacuum, and possibly add 
new strings. We first deal with 
holes created in the $j^{th}$ sea, suppressing some strings of length 
$a_{p}$, $p=1,2,\ldots$. 
The parameters $a_{p}$ have to be 
some of the $a_{\ell}$ parameters defining the representations of the
spin chain (i.e. $\frac{a_p-1}{2}\in \cN$).

\subsection{Valences and spin}

Let $\{\nu^{(j)}_{n}\}$ be the configuration of the vacuum.
The configuration corresponding to an excited state reads, 
for $n\in \cN$ and $1\leq j\leq N-1$,
\begin{equation}
\label{eq:tmug}
\wt\nu^{(j)}_{n}=\nu^{(j)}_{n}-\mu_{n}^{(j)}
\end{equation}
where the sets $\{\mu_{n}^{(j)}|n\in\cN\}$, $1\leq j\leq N-1$, 
 label the seas where perturbations to the vacuum configuration 
are introduced. Let us stress that $\nu^{(j)}_{n}$ as well as 
$\mu_{n}^{(j)}$ are a priori rational numbers (depending on $L$, the 
length of the chain) while $\wt\nu^{(j)}_{n}$ 
must be integers (they correspond to the number of $n$-string in the 
sea $j$). In the following, we will use 
\begin{equation}
\label{eq:v-mu}
v_{m}^{(j)} = 2\mu_{m}^{(j)}-\mu_{m}^{(j-1)}-\mu_{m}^{(j+1)}\qquad 
m\in\cN\,,\ j=1,\ldots,N-1
\end{equation}
with the convention $\mu_{m}^{(0)}=\mu_{m}^{(N)}=0$.

The corresponding valences are given by
\begin{equation}
\wt P^{(j)}_{n}=\wt\nu^{(j)}_{n}+\sum_{m\in\cN}\min(2n+1,2m+1)\,v_{m}^{(j)}
\,.
\end{equation}
They correspond to the number of allowed quantum numbers, while 
$\wt\nu^{(j)}_{n}$ denotes the number of parameters which are indeed 
used. 
Thus, for $n\in\cN$, 
\begin{equation}
\label{eq:D2}
\cD_n^{(j)}=\wt P^{(j)}_{n}-\wt\nu^{(j)}_{n}
=\sum_{m\in\cN}\min(2n+1,2m+1)\,v_{m}^{(j)}
\end{equation}
is the number of admissible quantum numbers 
$Q_{n,k}^{(j)}$ which are \underline{not} used.
They provide the number (for each sea and each length of string)
of free parameters attached to the excited 
state under consideration, i.e. the number of holes 
(formed by strings of length $2n+1$) in the $j^{th}$ sea. Thus, they 
must be postive.
Of course, some of the numbers $\cD_n^{(j)}$ may vanish. 
The spin of such a state is easily computed:
\begin{equation}
\label{eq:spins}
S_{j}=\sum_{m\in\cN}(2m+1)\,v_{m}^{(j)}=\cD_{n_\cL}^{(j)}\,.
\end{equation}
It is equal to the number of holes in the 
$j^{th}$ sea of the longest string.

In order to determine the first hole excitations of the model, 
one should look for the values of $v_m^{(j)}$ such that 
$\wt \nu_m^{(j)}$, $\cD_n^{(j)}$ and $S_j$ 
are positive integers and $\cD_n^{(j)}$ and/or $S_j$ have the smallest 
value. Such task seems impossible to solve on full general 
grounds, so that we will treat two slightly different problems: 
\textsl{(i)} 
Find excitations that exist whatever the type of chain
\textsl{(ii)} Find `small' excitations for a subclass of spin chain. 
The two following lemmas correspond to these two problems.
\begin{lemma}\label{lem1}
The hole excitations given by, for $1\leq\alpha\leq\cL$, 
\begin{eqnarray}
v_{n_\alpha}^{(j)}=\frac{\cJ_{\alpha,j}}{\cJ}= \frac{1}{\cJ}\sum_{\ell\in I_\alpha}
\delta_{j,j_{\ell}}\,,
\end{eqnarray}
where $\cJ_{\alpha,j}$ and  $\cJ$ are given in (\ref{def:cJ-aj}) and 
followings, exists for any $L_{0}$-regular spin chain. They have spin 
and hole numbers given by
(for $m\in\cN$
and $1\leq j\leq N-1$):
\begin{eqnarray}
\cD_{m}^{(j)} &=&\frac{1}{\cJ} \sum_{\ell=1}^{L_{0}}  
\delta_{j,j_{\ell}}\,\min(2m+1,a_{\ell}) \mb{and}
S_{j} \ =\ \frac{1}{\cJ} \sum_{\ell=1}^{L_{0}}  
\delta_{j,j_{\ell}}\,a_{\ell}\;.
\end{eqnarray}
\end{lemma}
These values generalize the ones obtained in \cite{fafa}, for the 
spin $s$ XXX model, and in \cite{anne}, for the $gl(N)$ spin chain built on 
fundamental representations (see also examples below). Moreover, 
since $v_{n_\alpha}^{(j)}$ are integers, equations (\ref{eq:D2}) and 
(\ref{eq:spins}) clearly show that
the number of holes $\cD_n^{(j)}$ and the spin $S_j$ for this state
are also integers. 
It leads, from the equations (\ref{eq:v-mu}),
to the following values (for $1\leq\alpha\leq\cL$
and $1\leq j\leq N-1$):
\begin{eqnarray}
\mu_{n_\alpha}^{(j)} &=&\frac{1}{N\, \cJ} \sum_{\ell\in I_\alpha}
\min(j,j_{\ell})(N-\max(j,j_{\ell}))\\
\wt\nu_{n_\alpha}^{(j)} &=& \frac{L-(L_{0}/\cJ)}{NL_{0}}\sum_{\ell\in I_\alpha}
\min(j,j_{\ell})(N-\max(j,j_{\ell}))\;.
\end{eqnarray}
We deduce that the corresponding excitation exists when the length 
of the chain, $L$, is such that $(L-(L_{0}/\cJ))$ is a multiple of $NL_{0}$. 
Such a length exists because $\cJ$ is also a divisor of $L_0$, see 
eq. (\ref{eq:L0cJ}).
The factor $1/\cJ$ may look weird at first sight.
However, it may be explained by remarking that the Bethe equations 
do not depend on the choice of the order of the representations\footnote{On 
the other hand, the order is crucial in the computation 
of conserved quantities deduced from the transfer matrix: their explicit form 
  may be completely different even though they have 
the same spectrum. In the same way, the order does matter for the 
calculation of Bethe vectors and correlation functions.} on 
the chain. Therefore, by reordering the representations, we may transform 
a $L_0$-regular spin chain to a $\frac{L_0}{\cJ}$-regular spin chain with
the same excitations. This may explain the presence of the denominator $\cJ$ 
in the above formulae.
These excitations correspond to the highest representation 
in the decomposition of the tensor product of the representations 
entering in the $\frac{L_0}{\cJ}$-regular chain. The Young tableau 
corresponding to this spin is the juxtaposition of the Young 
tableaux (in decreasing order of 
 $j_\ell$) of  all representations of the $\frac{L_0}{\cJ}$-regular 
chain (see third exemple below and Figure \ref{Youngtab-excited}).

\null

We now turn to the case \textsl{(ii)}. The aim is to encompass a wide 
class of spin chain (i.e. a wide class of possible representations on 
the chain), and in particular for $sl(2)$ spin chains, it 
will encompass \textsl{all} types of representation, 
provided\footnote{When $L_{0}=1$, one deals with spin $s$ $sl(2)$ spin 
chain, whose excitations are treated in lemma \ref{lem1}.} $L_{0}>1$. 
For $gl(N)$ spin 
chains, the representations entering the $L_{0}$-regular
chain will have rectangular 
Young tableaux of same height $j_{0}$ and arbitrary width $a_{\ell}$.
\begin{lemma}\label{lem2}
For $L_{0}$-regular spin chains with representations such that 
$j_{\ell}=j_{0}$, $\forall\,\ell$, and such that $(NL_{0})$ is even,
the following values define `small' 
hole excitations:
\begin{eqnarray}
v_{n_\alpha}^{(j)}=\frac{N}{2}\,\delta_{j,j_{0}}\,\eps_{\alpha} 
\mb{with} \eps_{\alpha}=\begin{cases}
 +1 \mb{when} \alpha=\cL,\cL-2,\ldots 
\\
-1 \mb{when} \alpha=\cL-1,\cL-3,\ldots 
\end{cases}
\end{eqnarray}
The hole numbers and spin are given by eqs (\ref{eq:D2}) and 
(\ref{eq:spins}).
\end{lemma}
It leads, from equations (\ref{eq:v-mu}),
to the following values (for $1\leq\alpha\leq\cL$
and $1\leq j\leq N-1$):
\begin{eqnarray}
\mu_{n_\alpha}^{(j)} &=&\frac{\eps_{\alpha}}{2} 
\min(j,j_{0})(N-\max(j,j_{0}))\\
\wt\nu_{n_\alpha}^{(j)} &=& \frac{2L-\eps_{\alpha}\,NL_{0}}{2NL_{0}}
\min(j,j_{0})(N-\max(j,j_{0}))\;.
\end{eqnarray}
We deduce that the corresponding excitation exists when the length 
of the chain, $L$, is such that $(L\pm \frac{NL_{0}}{2})$ is a multiple of $NL_{0}$.
It is obviously possible only when $NL_{0}$ is even. This is in 
particular ensured 
for any $sl(2)$ spin chains and also for alternating $gl(N)$ spin 
chains.

\begin{rmk}\label{rmk1}
Apart from the excitations introduced in lemmas \ref{lem1} and 
\ref{lem2}, there are many other possible excitations. In particular, 
any mutliple of these excitations is allowed: 
\begin{equation}
\wt\mu_{m}^{(j)} = k\,\mu_{m}^{(j)}\,,\qquad \forall m\in\cN\,,\ 
j=1,\ldots,N-1\,,\ k\in\ZZ_{+}
\end{equation}
where $\mu_{m}^{(j)}$ correspond to excitations of 
lemmas \ref{lem1} or \ref{lem2}. For instance, the spin 1 excitation 
of \cite{dVMN} corresponds to $k=2$ for an excitation of lemma \ref{lem2}.
\end{rmk}
\begin{rmk}\label{rmk2}
 For spin chains with a given length such that the 
vacuum exists (as determined in \cite{thermy}),  
 $\nu^{(j)}_{n}$ are integers, so that
excitations must have integer $\mu_n^{(j)}$. We will say that the 
corresponding excitations lie in the vacuum sector.
It is the point of view developed for example in \cite{Tak,fafa}.
\end{rmk}

\subsubsection*{Examples}

\quad\textit{1.} We consider a $gl(N)$ spin chain where
 all sites carry the fundamental representation $\underline{N}$. 
Since the vacuum 
 state is built on real strings only, it is the only type of strings 
one can suppress. Then, 
 \begin{equation}
\wt\nu^{(j)}_{0}=\frac{L(N-j)}{N}-\mu^{(j)}\mb{;}
\wt\nu^{(j)}_{n}=0\,,\ n>0
\end{equation}
and we get
\begin{equation}
S_{j}=\cD^{(j)}_0=2\mu^{(j)}-\mu^{(j-1)}-\mu^{(j+1)}\;,
\end{equation}
in accordance with the results given above.

The excitation carrying the
$sl(N)$ fundamental representation $\vec{S}=(1,0,0,\ldots,0)$ 
corresponds to the values given in lemma \ref{lem1}:
\begin{equation}
    \mu^{(j)}=\frac{N-j}{N}\,.
\end{equation}
The number of quasi-particles is
$\cD^{(j)}_n=\delta_{n,0}\,\delta_{j,1}$, as expected.
 This `first' excitation  lies in a spin chain with length such
that $L-1$ is a multiple of $N$. We remind that the vacuum is lying
in a spin chain with $L$ multiple of $N$.
\\

\textit{2.} For a $sl(2)$ spin chain with a spin $s$ representation on each
site, the vacuum is built on $s-\half$ strings only. Thus, we get
 \begin{equation}
\wt\nu_{s-\half} = \frac{L}{2}-\mu\mb{and} \wt\nu_{n} = 0\,,\ \forall\,
n\in\half\ZZ\,,\ n\neq s-\half\,.
\end{equation}
The resulting excited state has spin\footnote{The factor $\half$ is 
implemented to make contact with the usual spin used for $sl(2)$.} 
$\frac{S}{2}= 2s\,\mu$ and a number 
$\cD_{n}=4s\,\mu\,\delta_{n,s-\half}$ of spin $\half$ `elementary particles' 
\cite{Tak}. 
The hole excitation of lemma \ref{lem1} corresponds to $\mu=\half$ (i.e. $v=1$), 
so that one can get a state with spin $s$. 
This spin 
$s$ excitation is obtained when $L$ is odd, while the vacuum has $L$
even.\\

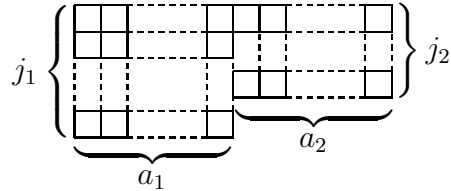
\begin{figure}[ht]
\begin{displaymath}
  \raisebox{22pt}{$j_1 \left\{
  \rule{0pt}{30pt}
  \right.$}
  \begin{picture}(60,60)
    \put(0,30){\line(1,0){20}}\put(50,30){\line(1,0){10}}
    \put(0,50){\line(1,0){20}}\put(50,50){\line(1,0){10}}
    \put(0,40){\line(1,0){20}}\put(50,40){\line(1,0){10}}
    \put(0,50){\line(0,-1){20}}\put(50,50){\line(0,-1){20}}
    \put(10,50){\line(0,-1){20}}\put(60,50){\line(0,-1){20}}
    \put(20,50){\line(0,-1){20}}
    \put(0,10){\line(1,0){20}}\put(50,10){\line(1,0){10}}
    \put(0,0){\line(1,0){20}}\put(50,0){\line(1,0){10}}
    \put(0,10){\line(0,-1){10}}\put(50,10){\line(0,-1){10}}
    \put(10,10){\line(0,-1){10}}\put(60,10){\line(0,-1){10}}
    \put(20,10){\line(0,-1){10}}
    \multiput(20,50)(4,0){8}{\line(1,0){2}}
    \multiput(20,40)(4,0){8}{\line(1,0){2}}
    \multiput(20,30)(4,0){8}{\line(1,0){2}}
    \multiput(20,10)(4,0){8}{\line(1,0){2}}
    \multiput(20,0)(4,0){8}{\line(1,0){2}}
    \multiput(0,10)(0,4){5}{\line(0,1){2}}
    \multiput(10,10)(0,4){5}{\line(0,1){2}}
    \multiput(20,10)(0,4){5}{\line(0,1){2}}
    \multiput(50,10)(0,4){5}{\line(0,1){2}}
    \multiput(60,10)(0,4){5}{\line(0,1){2}}
    \put(0,-3){$\underbrace{~~~~~~~~~~~~~~~}_{\displaystyle a_1}$}
  \end{picture}
\begin{picture}(60,50)
    \put(0,50){\line(1,0){20}}\put(50,50){\line(1,0){10}}
    \put(0,40){\line(1,0){20}}
    \put(50,40){\line(1,0){10}}\put(50,50){\line(0,-1){10}}
    \put(10,50){\line(0,-1){10}}\put(60,50){\line(0,-1){10}}
    \put(20,50){\line(0,-1){10}}
    \put(0,15){\line(1,0){20}}\put(50,15){\line(1,0){10}}
    \put(0,25){\line(1,0){20}}
    \put(50,25){\line(1,0){10}}
    \put(10,25){\line(0,-1){10}}\put(50,25){\line(0,-1){10}}
    \put(10,25){\line(0,-1){10}}\put(60,25){\line(0,-1){10}}
    \put(20,25){\line(0,-1){10}}\put(0,25){\line(0,-1){10}}
    \multiput(20,50)(4,0){8}{\line(1,0){2}}
    \multiput(20,40)(4,0){8}{\line(1,0){2}}
    \multiput(20,15)(4,0){8}{\line(1,0){2}}
    \multiput(20,25)(4,0){8}{\line(1,0){2}}
    \multiput(10,25)(0,4){5}{\line(0,1){2}}
    \multiput(20,25)(0,4){5}{\line(0,1){2}}
    \multiput(50,25)(0,4){5}{\line(0,1){2}}
    \multiput(60,25)(0,4){5}{\line(0,1){2}}
    \put(1,11){$\underbrace{~~~~~~~~~~~~~~~}_{\displaystyle a_2}$}
  \end{picture}
    \raisebox{30pt}{$ \left.
  \rule{0pt}{20pt}
  \right\}j_2$}
\end{displaymath}
    \caption{Young tableau of the hole excitation for an alternating spin chain 
    $(a_1,j_1)$-$(a_2,j_2)$ \label{Youngtab-excited}}
\end{figure}

\textit{3.}
If one considers an alternating spin chain, where the sites $2\ell+1$ 
carry an $sl(N)$ representation defined by $(a_{1},j_{1})$ and the sites $2\ell$ 
carry a representation $(a_{2},j_{2})$. We
suppose that $a_{1}< a_{2}$, so that $a_{\ell}=2n_{\ell}+1$. Hole 
excitations have configurations:
\begin{eqnarray}
\wt\nu^{(j)}_{n}&=&\nu^{(j)}_{n}\mb{for} n\neq n_{1},n_{2}\\
\wt\nu^{(j)}_{n_{k}}&=&\nu^{(j)}_{n_{k}}-\,\mu_{n_{k}}^{(j)}\mb{for} k=1,2
\end{eqnarray}
The spin is
\begin{equation}
 S_{j} = a_{1}\,(2\mu_{n_{1}}^{(j)}-\mu_{n_{1}}^{(j-1)}-\mu_{n_{1}}^{(j+1)})
 +a_{2}\,(2\mu_{n_{2}}^{(j)}-\mu_{n_{2}}^{(j-1)}-\mu_{n_{2}}^{(j+1)})
\end{equation}
The excitations of lemma \ref{lem1} are given by 
\begin{equation}
\begin{cases}
\displaystyle \mu_{n_{k}}^{(j)}=\frac{j(N-j_{k})}{N}\,  
& j\leq j_{k} \\[1.2ex]
\displaystyle \mu_{n_{k}}^{(j)}=\frac{j_{k}(N-j)}{N}\,  & j_{k}\leq j 
\end{cases} \qquad k=1,2\,.
\end{equation}
They lead to non-vanishing hole numbers
\begin{eqnarray}
\cD_{n_{1}}^{(j)} &=& (2n_{1}+1)(\delta_{j,j_{1}}+\delta_{j,j_{2}})\\
\cD_{n_{2}}^{(j)} &=& (2n_{1}+1)\delta_{j,j_{1}}
+(2n_{2}+1)\delta_{j,j_{2}}\,.
\end{eqnarray}
The spin $S=(S_{1},\ldots,S_{N-1})$ is given by 
\begin{equation}
\displaystyle S_{j}=0 \mb{for} j\neq j_{1},j_{2}\mb{and}
\displaystyle S_{j_{k}}=a_{k}\mb{for} k=1,2
\end{equation}
corresponding to a representation with Young tableau given 
in Figure \ref{Youngtab-excited} (drawn for $j_1>j_2$).

If we suppose furthermore that $j_{1}=j_{2}=j_{0}$, one can consider
 excitations as in lemma \ref{lem2}:
\begin{equation}
\begin{cases}
\displaystyle \mu_{n_{k}}^{(j)}=(-1)^k\,\frac{j(N-j_{0})}{2}\,  
& j\leq j_{0} \\[1.2ex]
\displaystyle \mu_{n_{k}}^{(j)}=(-1)^k\,\frac{j_{0}(N-j)}{2}\,  & j_{0}\leq j 
\end{cases} \qquad k=1,2
\end{equation}
They lead to a unique non vanishing hole number 
$\cD_{n_{2}}^{(j)}=N(n_{2}-n_{1})\,\delta_{j,j_{0}}$
and have a spin $S=(S_{1},\ldots,S_{N-1})$ with 
\begin{equation}
\displaystyle S_{j}=0 \mb{for} j\neq j_{0}\mb{and}
\displaystyle S_{j_{0}}=N(n_{2}-n_{1})\,.
\end{equation}

\subsection{Thermodynamical limit}

For $m\in \cN$, the quantum numbers for an excited state are
\begin{equation}
  \wt{Q}_{m,k}^{(j)}=k-\half(\wt{\nu}_{m}^{(j)}+1)
+\half \sum_{d=1}^{\cD^{(j)}_m}\text{sign}(k-k_{m,d}^{(j)})\, \quad
k=1,\ldots,\wt{\nu}_{m}^{(j)}\;,
\label{eq:Qmk-exc1}
\end{equation}
where we have introduced the \textit{sign} function
\begin{equation}
\text{sign}(j)=\begin{cases}
1& j\geq0\\
-1&j<0
\end{cases}\;.
\end{equation}
The numbers $k_{m,d}^{(j)}$ are integers characterizing the holes.
They must fulfill the constraint
\begin{equation}
1\leq k_{m,d}^{(j)}\leq \wt{\nu}_{m}^{(j)}\;.
\end{equation}

As before, we take the thermodynamical limit of the Bethe ansatz
 equation and we obtain (see e.g. \cite{fafa}):
\begin{eqnarray}
&&\sum_{n\in\cN}\left\{
\int_{-\infty}^{\infty} 
d\lambda \,\Big({\fs}_{n}^{(j-1)}(\lambda)+
\fs_{n}^{(j+1)}(\lambda)\Big)\,
{\Psi_{-1}^{(m,n)}}(\lambda_{0}-\lambda)
+\int_{-\infty}^{\infty} d\lambda \,\fs_{n}^{(j)}(\lambda)\, 
{\Psi_{2}^{(m,n)}}(\lambda_{0}-\lambda)
\right\}\nonu
&&\qquad
=-2\pi\,\left(\fs_{m}^{(j)}(\lambda_0)+\frac{1}{L}
\sum_{d=1}^{\cD^{(j)}_m} \delta(\lambda_0-\wt\lambda_{m,d}^{(j)})\right)
+\frac{1}{L_{0}}\sum_{\alpha=1}^{\cL}\Big(\sum_{\ell\in I_\alpha}\delta_{j,j_{l}}\Big)\,
\Psi_{\bar a_{\alpha}}^{(m)}(\lambda_{0})\qquad\qquad\\
&&\qquad \forall \lambda_{0}\in\;
 ]-\infty\,,\,\infty[\ ,\ \forall
j=1,\ldots,N-1\ ,\ \forall m\in\cN\nonumber
\end{eqnarray}
where $\fs_{m}^{(j)}(\lambda)$ is the density of $(2m+1)$-strings  
in the sea $j$ for the considered excited state. The 
parameter $\wt\lambda_{m,d}^{(j)}$ is the image of $k_{m,d}^{(j)}/L$
in the limit $L\to\infty$.

Performing a Fourier transform, the Bethe ansatz equations become
\begin{equation}
2\pi\wh\Psi(p)\, \wh{S}(p)=\Lambda(p)-\frac{1}{L}\Delta(p)\;,
\end{equation}
where $\wh\Psi(p)$, $\Lambda(p)$ are given in section \ref{sec:sum}
and
\begin{eqnarray}
\Delta(p)&=&\sum_{j=1}^{N-1}\sum_{\alpha=1}^{\cL}\ 
\sum_{d=1}^{\cD_{n_\alpha}^{(j)}}\ \exp(ip\, \wt\lambda_{n_\alpha,d}^{(j)})
\ e_j^{(N-1)}\otimes e_\alpha^{(\cL)}
\\
\wh{S}(p)&=&
\sum_{j=1}^{N-1}\sum_{\alpha=1}^{\cL}\ 
\wh{\fs}_{n_\alpha}^{(j)}(p)
\;e_j^{(N-1)}\otimes e_\alpha^{(\cL)}\; .
\end{eqnarray}
By linearity, the solutions take the 
following form
\begin{equation}
\label{eq:sL}
\wh S(p)=\wh \Sigma(p)+\frac{1}{L}\wh R(p)
\end{equation}
where $\wh \Sigma(p)$ gathers the densities for the vacuum (\ref{eq:SigLam}) and
 $\displaystyle \wh R(p)=\sum_{j=1}^{N-1}\sum_{\alpha=1}^{\cL}\ 
\wh \rho_{n_\alpha}^{(j)}(p)\ 
\;e_j^{(N-1)}\otimes e_\alpha^{(\cL)}$ corresponds to the densities 
of holes. It is solution of the following linear 
system
\begin{equation}
2\pi\wh\Psi(p)\, \wh{R}(p)=-\Delta(p)\;.
\end{equation}
Inverting the matrix $\wh\Psi(p)$, we get the solution
\begin{equation}
\wh R(p)=\frac{1}{2\pi}\,\Big(\cA(p)^{-1}
\otimes \big(\wh\Psi_{-1}(p)\big)^{-1}\Big)\;\Delta(p)
\label{BAE-TF-inv-ex}
\end{equation}
where the inverse of the matrix $\cA(p)$ is given by (see e.g. \cite{sutherland})
\begin{equation}
(\cA(p)^{-1})_{ij}=
\frac{\displaystyle\sinh\Big((N-\max(i,j))\frac{\vert p\vert}{2}\Big) \;
\sinh\Big(\min(i,j)\frac{\vert p\vert}{2}\Big)} 
{\displaystyle\sinh\Big(N\frac{\vert p\vert}{2}\Big) \; 
\sinh\Big(\frac{\vert p\vert}{2}\Big)}\;.
\label{eq:psi}
\end{equation}
The inverse of the $\cL\times \cL$ matrix $\wh\Psi_{-1}(p)$ is provided by 
the following lemma (we remind that we ordered $n_1<n_2<...<n_\cL$).
\begin{lemma}\label{lem:inv}
The inverse of $\wh \Psi_{-1}(p)$ is a symmetric tridiagonal matrix.
Its diagonal elements are given by, for $1\leq\alpha\leq\cL$
\begin{eqnarray}
\Big(\wh \Psi_{-1}(p)\Big)^{-1}_{\alpha\alpha}=
-\ 
\frac{\sinh\big(\frac{\vert p\vert }{2}\big)\ 
\sinh\big(\vert p\vert \;(n_{\alpha+1}-n_{\alpha-1})\big)}
{\sinh\big(\vert p\vert \;(n_{\alpha}-n_{\alpha-1})\big)
\sinh\big(\vert p\vert \;(n_{\alpha+1}-n_\alpha)\big)}
\end{eqnarray}  
where, by convention, we set $n_0=-\frac{1}{2}$ and 
$n_{\cL+1}=+\infty$. We also used the limit
$\frac{\sinh(\infty-a)}{\sinh(\infty-b)}=\exp(b-a)$. 
The non-vanishing non-diagonal elements read, for $1\leq \alpha< \cL-1$,
\begin{equation}
\Big(\wh \Psi_{-1}(p)\Big)^{-1}_{\alpha,\alpha+1}=
\frac{\sinh\big(\frac{\vert p\vert}{2}\big)}
{\sinh\big(\vert p\vert \;(n_{\alpha+1}-n_\alpha)\big)}
\ =\ \Big(\wh \Psi_{-1}(p)\Big)^{-1}_{\alpha+1,\alpha}
\end{equation}
In the particular case $\cL=1$, this matrix reduces to a number: 
\begin{equation}
\Big(\wh \Psi_{-1}(p)\Big)^{-1}=-\sinh\big(\frac{\vert p\vert}{2}\big)\
\frac{e^{\big(\vert p\vert\; (n_{1}+\frac{1}{2})\big)}}
{\sinh\big(\vert p\vert\;(n_{1}+\frac{1}{2})\big)}
\end{equation}
\end{lemma}
\prf Direct computation on the product $\Big(\wh 
\Psi_{-1}(p)\Big)^{-1}\ \wh \Psi_{-1}(p)$.
\finprf
Let us introduce useful functions, for $1\leq \alpha\leq \cL$,
\begin{equation}
\label{eq:h}
\wh{h}^{(\alpha)}_{j,k}(p)=-\ \frac{1}{2\pi}\ 
\frac{\displaystyle\sinh\Big(\big(N-\max(j,k)\big)\frac{\vert p\vert}{2}\Big) \;
\sinh\Big(\min(j,k)\frac{\vert p\vert}{2}\Big)} 
{\displaystyle
\sinh\Big(N\frac{\vert p\vert}{2}\Big)\
\sinh\Big((n_{\alpha+1}-n_{\alpha})\vert p\vert\Big) }
\end{equation}
and
\begin{equation}
\label{eq:g}
\wh{g}^{(\alpha)}_{j,k}(p)=\wh{h}^{(\alpha)}_{j,k}(p)
\frac{\displaystyle
\sinh\Big((n_{\alpha+1}-n_{\alpha-1})\;\vert p\vert\Big)} 
{\displaystyle
\sinh\Big((n_{\alpha}-n_{\alpha-1})\;\vert p\vert\Big) }\ 
\end{equation}
where the convention $n_0=-1/2$ and $n_{\cL+1}=\infty$ has been applied. 
\begin{theorem} 
\label{thm:density-excited}
Let us consider a $L_{0}$-regular spin chain based on $gl(N)$. 
For the excited states corresponding to holes in the filled seas, 
the densities are, for $1\le j \le N-1$ and $1\leq \alpha\leq \cL$,
\begin{equation}
\fs^{(j)}_{n_\alpha}(\lambda)=\sigma^{(j)}_{n_\alpha}(\lambda)
+\frac{1}{L}\rho^{(j)}_{n_\alpha}(\lambda)
\label{eq:s}
\end{equation}
where $\sigma^{(j)}_{n_\alpha}(\lambda)$ are the densities 
(\ref{eq:densitemagique}) for the vacuum and 
the corrections at order $1/L$ are
\begin{eqnarray}
\rho^{(j)}_{n_\alpha}(\lambda)&=&
\sum_{k=1}^{N-1}\left(
-\sum_{d=1}^{\cD_{n_{\alpha-1}}^{(k)}}
h^{(\alpha-1)}_{jk}(\lambda-\wt \lambda_{n_{\alpha-1},d}^{(k)})
+\sum_{d=1}^{\cD_{n_\alpha}^{(k)}}
g^{(\alpha)}_{jk}(\lambda-\wt \lambda_{n_\alpha,d}^{(k)})
-\sum_{d=1}^{\cD_{n_{\alpha+1}}^{(k)}}
h^{(\alpha)}_{jk}(\lambda-\wt \lambda_{n_{\alpha+1},d}^{(k)})
\right)\;.\quad
\label{eq:densite-correction}
\end{eqnarray}
The functions $h^{(\alpha)}_{jk}(\lambda)$ and 
$g^{(\alpha)}_{jk}(\lambda)$ are
 the inverse Fourier transform of (\ref{eq:h}) and (\ref{eq:g}) respectively. 
By convention, we used $\cD_{n_{0}}^{(k)}=\cD_{-1/2}^{(k)}=0$ and 
$\cD_{n_{\cL+1}}^{(k)}=\cD_{\infty}^{(k)}=0$.

The corrections to the densities, for the particular case $\cL=1$, reduce to
\begin{eqnarray}
\rho^{(j)}_{n_1}(\lambda)&=&
\sum_{k=1}^{N-1}\ \sum_{d=1}^{\cD_{n_1}^{(k)}}
g_{jk}(\lambda-\wt \lambda_{n_1,d}^{(k)})
\label{eq:densite-correctionL1}
\end{eqnarray}
where $g_{jk}(\lambda)=g_{jk}^{(1)}(\lambda)$ 
(for $\cL=1,\;n_0=-1/2,\;n_2=\infty$).
\end{theorem}
\prf 
We give the proof only for $\alpha=1$. The other density 
corrections are obtained similarly.
Projecting the linear system (\ref{BAE-TF-inv-ex}) on its first component 
and using the lemma \ref{lem:inv}, we get
\begin{equation}
\wh \rho_{n_1}^{(j)}(p)=
\sum_{k=1}^{N-1}\left(\sum_{d=1}^{\cD_{n_1}^{(k)}}
\wh{g}^{(1)}_{jk}(p)\exp(ip\wt\lambda_{n_1,d}^{(k)})
-\sum_{d=1}^{\cD_{n_2}^{(k)}}
\wh h^{(1)}_{jk}(p)\exp(ip\wt\lambda_{n_2,d}^{(k)})
\right)
\end{equation}
Performing an inverse Fourier transform and using the conventions 
$n_0=-\half$ and $\cD_{n_{0}}^{(k)}=0$, we get the result 
(\ref{eq:densite-correction}) for $\alpha=1$.
\finprf
Let us remark that the explicit expressions of 
$h^{\alpha}_{jk}(\lambda)$ and $g^{\alpha}_{jk}(\lambda)$ 
are not needed, since physical quantities 
are computed via their Fourier transform (see for instance section \ref{sec:energy}).

\subsection{Examples}

\quad\textit{1.} For a $gl(N)$ fundamental spin chain, the corrections to the
real root densities are
\begin{eqnarray}\label{eq:densite-correctionL1-ex1}
\wh\rho^{(j)}_{0}(p)=-\ \frac{1}{2\pi}\ 
\sum_{k=1}^{N-1}\ \sum_{d=1}^{\cD_{0}^{(k)}}\
\frac{\displaystyle\sinh\Big((N-\max(j,k))\frac{\vert p\vert}{2}\Big) \;
\sinh\Big(\min(j,k)\frac{\vert p\vert}{2}\Big)} 
{\displaystyle
\sinh\Big(N\frac{\vert p\vert}{2}\Big)\
\sinh\Big(\frac{\vert p\vert}{2}\Big) }\ 
e^{\frac{\vert p\vert}{2}+ip\wt\lambda_{d}^{(k)}}
\end{eqnarray}
where we noted $\wt\lambda^{(k)}_{d}$ for $\wt\lambda^{(k)}_{0,d}$.

\textit{2.} For a $gl(2)$ spin $s$ chain, the corrections to the
root densities of length $2s$ are
\begin{eqnarray}\label{eq:densite-correctionL1-ex2}
\wh\rho_{s-1/2}(p)&=&-\ \frac{1}{2\pi}\ 
\sum_{d=1}^{\cD_{s-1/2}}\
\frac{\displaystyle\sinh^2\Big(\frac{\vert p\vert}{2}\Big)} 
{\displaystyle
\sinh(\vert p\vert)\
\sinh(s\vert p\vert) }\ 
e^{s\vert p\vert+ip\wt\lambda_{d}}
\end{eqnarray}
where $\wt\lambda_{d}\equiv\wt\lambda_{s-\half,d}^{(1)}$.

\textit{3.} For a $gl(N)$ alternating spin chain with $n_1<n_2$, we get
\begin{eqnarray}\label{eq:densite-correctionL1-ex3}
\wh\rho^{(j)}_{n_1}(p)&=&-\ \frac{1}{2\pi}\ 
\sum_{k=1}^{N-1}\ \ 
\frac{\displaystyle\sinh\Big((N-\max(j,k))\frac{\vert p\vert}{2}\Big) \;
\sinh\Big(\min(j,k)\frac{\vert p\vert}{2}\Big)
} 
{\displaystyle
\sinh\Big(N\frac{\vert p\vert}{2}\Big)\
\sinh\Big((n_2-n_1)\,\vert p\vert\Big)
 }\nonumber\\
&&\hspace{1.5cm}\times
\left\{
\frac{\sinh((n_2+\half)\vert p \vert)}{\sinh((n_1+\half)\vert p \vert)}\
\sum_{d=1}^{\cD_{n_1}^{(k)}}\ e^{ip\wt\lambda_{n_1,d}^{(k)}}
-\sum_{d=1}^{\cD_{n_2}^{(k)}}\ e^{ip\wt\lambda_{n_2,d}^{(k)}}
\right\}
\end{eqnarray}
and
\begin{eqnarray}\label{eq:densite-correctionL1-ex3bis}
\wh\rho^{(j)}_{n_2}(p)&=&-\ \frac{1}{2\pi}\ 
\sum_{k=1}^{N-1}\ \ 
\frac{\displaystyle\sinh\Big((N-\max(j,k))\frac{\vert p\vert}{2}\Big) \;
\sinh\Big(\min(j,k)\frac{\vert p\vert}{2}\Big)
} 
{\displaystyle
\sinh\Big(N\frac{\vert p\vert}{2}\Big)\
\sinh\Big((n_2-n_1)\,\vert p\vert\Big)
 }\ 
\nonumber\\
&&\hspace{1.5cm}\times\left\{-
\ \sum_{d=1}^{\cD_{n_1}^{(k)}}\ e^{ip\wt\lambda_{n_1,d}^{(k)}}
+\exp((n_2-n_1)\vert p \vert)
\sum_{d=1}^{\cD_{n_2}^{(k)}}\ e^{ip\wt\lambda_{n_2,d}^{(k)}}
\right\}
\end{eqnarray}

\section{General excited states\label{sec:gh}}

We now turn to a more general excited state. It corresponds 
to the case where holes are created in the filled seas and  
 new strings added in other seas. 

\subsection{Valences and spins}

The configuration of an excited state is characterized by
\begin{equation}
\wt\nu^{(j)}_{n}=\nu^{(j)}_{n}-\mu_{n}^{(j)}\mb{for} n\in \cN
\mb{and} \wt\nu^{(j)}_{n}\geq 0 \mb{for} n\in\overline\cN\,,
\end{equation}
where we kept the notation $\nu^{(j)}_{n}$, $n\in\cN$, for the vacuum 
configuration. The set $\bcN$ (complementary to $\cN$) has been 
introduced in (\ref{def:barN}). Obviously, 
if $\wt\nu^{(j)}_{n}= 0,$ 
 $\forall n \in \overline\cN$, we recover the previous case with  
 holes only. 

The corresponding valences and spins are given by, for $n\in\half\ZZ_{\geq 0}$,
\begin{eqnarray}
\wt P_n^{(j)}&=&\wt\nu^{(j)}_{n}
+\sump_{m\in\cN}\min(2n+1,2m+1)\,(2\mu_{m}^{(j)}-\mu_{m}^{(j-1)}-\mu_{m}^{(j+1)})\nonumber\\
&&\hspace{1cm}-\sump_{m\in\overline\cN}\min(2n+1,2m+1)\,(2\wt\nu_{m}^{(j)}
-\wt\nu_{m}^{(j-1)}-\wt\nu_{m}^{(j+1)})\\
S_j&=&\sump_{m\in\cN}(2m+1)\,(2\mu_{m}^{(j)}-\mu_{m}^{(j-1)}-\mu_{m}^{(j+1)})
-\sump_{m\in\overline\cN}(2m+1)\,(2\wt\nu_{m}^{(j)}
-\wt\nu_{m}^{(j-1)}-\wt\nu_{m}^{(j+1)})\;.
\end{eqnarray} 
Obviously, demanding the valences and the spin to be positive imposes 
contraints between the number of holes,
$\cD_m^{(j)}=\wt P_m^{(j)}-\wt\nu^{(j)}_{m}$ ($m\in\cN$), and the number of new 
strings, $\wt\nu^{(j)}_{n}$ ($n\in\overline \cN$). 
 We  study in more details particular excitations in section 
 \ref{sec:spin-min}.
\\
The new strings of length $2n+1$ with $n\in \bcN$ are characterized by 
quantum numbers $\{Q_{n,k}^{(j)}\ |\ 1\leq k\leq \wt \nu_n^{(j)}\}$ and by the Bethe roots
$\{\lambda_{n,k}^{(j)}\ |\ 1\leq k\leq \wt \nu_n^{(j)}\}$ (these 
sets may be empty). These quantum numbers are bounded, namely, for $n\in\overline\cN$,
\begin{equation}
\frac{1-\wt P_n^{(j)}}{2}\leq Q_{n,k}^{(j)} \leq \frac{\wt P_n^{(j)}-1}{2}\;.
\end{equation}

Let us emphasize the difference between 
$\lambda_{n,k}^{(j)}$ ($n\in\overline\cN$), Bethe roots 
for the new strings (the ones added to the vacuum configuration), and 
$\wt\lambda_{n,k}^{(j)}$ ($n\in\cN$), missing Bethe roots in 
the filled seas (holes in the vacuum configuration). 

\subsection{Densities for a general excited state\label{sec:glExcit}}

In this section, we compute the densities for the general excited states in the 
thermodynamical limit. This is done in two steps: first, 
the Bethe equations for $n\in\cN$ provide the 
correction to the densities in terms of the new Bethe roots $\lambda_{n,k}^{(j)}$, 
$n\in\overline\cN$ (see theorem \ref{th:cj}). Second, the Bethe equations for 
$n\in\overline\cN$ couple these new Bethe roots with the holes in the filled seas 
(see theorem \ref{th:constraint}). 
These relations depend on the choice of the quantum numbers 
$\{Q_{n,k}^{(j)}\ |\ n\in\overline\cN,1\leq k\leq \wt \nu_n^{(j)}\}$.

Using again linearity of the Bethe equations, the densities 
for these excited states rewrite:
\begin{eqnarray}
\ft^{(j)}_{n}(\lambda) &=&
\fs^{(j)}_{n}(\lambda)+\frac{1}{L}\fc^{(j)}_{n}(\lambda)\;,
\quad n\in \cN\;,\label{eq:correction}\\
\ft^{(j)}_{m}(\lambda) &=&
\frac{1}{L}\,\sum_{\ell=1}^{\nu^{(j)}_{m}} 
\delta(\lambda-\lambda^{(j)}_{\ell,m})\;,
\quad m\in \bcN\;,\label{eq:correction2}
\end{eqnarray}
where 
$\fs^{(j)}_{n}(\lambda)=\sigma^{(j)}_{n}(\lambda)+\frac{1}{L}\rho^{(j)}_{n}(\lambda)$ 
are the densities when only holes are created in the vacuum 
configuration, as given in theorem \ref{thm:density-excited}, while
$\fc^{(j)}_{n}(\lambda)$ are the corrections due to the new added 
strings. We have assumed that the number of these new strings remains 
finite even in the limit $L\to\infty$, hence the form 
(\ref{eq:correction2}).

From the Bethe equations for $m\in\cN$ and performing a 
calculation similar to the one of section \ref{sec:holExcit}, one shows that 
$\fc^{(j)}_{n}(\lambda)$ must satisfy the following equations 
\begin{eqnarray}
&&\hspace{-1cm}\sump_{n\in\cN}\left\{
\int_{-\infty}^{\infty} 
d\lambda \,\Big({\fc}_{n}^{(j-1)}(\lambda)+
\fc_{n}^{(j+1)}(\lambda)\Big)\,
{\Psi_{-1}^{(m,n)}}(\lambda_{0}-\lambda)
+\int_{-\infty}^{\infty} d\lambda \,\fc_{n}^{(j)}(\lambda)\, 
{\Psi_{2}^{(m,n)}}(\lambda_{0}-\lambda)
\right\}+2\pi\,\fc_{m}^{(j)}(\lambda_0)\nonu
&&\label{eq:betc}\hspace{-1cm}+
\sump_{n\in\overline\cN}\left\{\sum_{\ell=1}^{\wt\nu^{(j-1)}_{n}}
\Psi_{-1}^{(m,n)}(\lambda_{0}-\lambda_{n,\ell}^{(j-1)})
+\sum_{\ell=1}^{\wt\nu^{(j)}_{n}}
\Psi_{2}^{(m,n)}(\lambda_{0}-\lambda_{n,\ell}^{(j)})
+\sum_{\ell=1}^{\wt\nu^{(j+1)}_{n}}
\Psi_{-1}^{(m,n)}(\lambda_{0}-\lambda_{n,\ell}^{(j+1)})
\right\} =0
\\
&&\qquad \forall \lambda_{0}\in\;
 ]-\infty\,,\,\infty[\ ,\ \forall
j=1,\ldots,N-1\ ,\ \forall m\in\cN\nonumber
\end{eqnarray}

Before solving these equations, we introduce 
the maps $\gamma^\pm$:
\begin{equation}
m\in\bcN\ \mapsto\ \begin{cases}
\gamma^{-}(m)\,=\, n_{\alpha}\in \cN\cup\{n_0=-\half,n_{\cL+1}=\infty\}
\\[1.2ex]
\gamma^{+}(m)\,=\, n_{\alpha+1}\in \cN\cup\{n_0=-\half,n_{\cL+1}=\infty\}
\end{cases}
\mbox{with }\ n_{\alpha}<m<n_{\alpha+1}
\;.
\end{equation}
In words, the functions $\gamma^\pm$ associate to any 
element in $\overline \cN$, the two closest numbers 
belonging to $\cN\cup\{-\half,\infty\}$.

We also need the following functions, for $m\in \overline \cN$,
\begin{equation}
\wh\omega_m^\pm(p)=
\frac{\sinh\big( p\, \vert m-\gamma^\pm(m)\vert\big)}
{\sinh\big( p \,(\gamma^+(m)-\gamma^-(m))\big)}\;.
\end{equation}
Their inverse Fourier transform is 
\begin{equation}
\omega^\pm_m(\lambda)=\frac{\pi}{\gamma^+(m)-\gamma^-(m)}\ 
\frac{\displaystyle
\sin\left(\frac{\pi\,\vert m-\gamma^\pm(m)\vert}{\gamma^+(m)-\gamma^-(m)}\right)}
{\displaystyle
\cosh\left(\frac{\pi\,\lambda}{\gamma^+(m)-\gamma^-(m)}\right)+
\cos\left(\frac{\pi\,\vert m-\gamma^\pm(m)\vert}{\gamma^+(m)-\gamma^-(m)}\right)}
\end{equation}
Let us remark that, for $m\in\overline\cN$ finite and such that $\gamma^+(m)=\infty$, 
these functions reduce to
\begin{equation}
\omega^+_m(\lambda)=
\frac{2\,(m-\gamma^-(m))}{\lambda^2+ (m-\gamma^-(m))^2}
\mb{and} \omega^-_m(\lambda)=0
\end{equation}
Now, we can give the corrections to the densities for a 
general excited state.
\begin{theorem}\label{th:cj}
The densities for a general excited state are given by 
(\ref{eq:correction2}) and by
\begin{equation}
\ft^{(j)}_{n}(\lambda)=\fs^{(j)}_{n}(\lambda)
+\frac{1}{L}\fc^{(j)}_{n}(\lambda)\;,\quad n\in\cN
\label{eq:t}
\end{equation}
where $\fs^{(j)}_{n}(\lambda)$ are given in theorem \ref{thm:density-excited}. 
The corrections $\fc^{(j)}_{n}(\lambda)$ are given by, 
for $1\leq \alpha\leq \cL$ and $1\leq j\leq N-1$,
\begin{equation}
\label{eq:c}
\fc_{n_\alpha}^{(j)}(\lambda)=-\frac{1}{2\pi}\left(
\sump_{n_{\alpha-1}<m<n_{\alpha}}
\sum_{\ell=1}^{\wt \nu_m^{(j)}}
\omega_{m}^-(\lambda-\lambda_{m,\ell}^{(j)})\ 
+\sump_{n_{\alpha}<m<n_{\alpha+1}}
\sum_{\ell=1}^{\wt \nu_m^{(j)}}
\omega_{m}^+(\lambda-\lambda_{m,\ell}^{(j)})\right)\quad
\end{equation}
 We remind that
$n_0=-1/2$ and $n_{\cL+1}=\infty$. By convention, we also set
$\omega^-_0(\lambda)=0$.
\end{theorem}
\prf
Performing a Fourier transform of equation (\ref{eq:betc}) and 
inverting the linear system, we get, for $1\leq \alpha \leq \cL$,
\begin{equation}
\wh{\fc}_{n_\alpha}^{(j)}(p)=-\frac{1}{2\pi}
\sump_{m\in\overline\cN}\ \sum_{\ell=1}^{\wt \nu_m^{(j)}}
\ \exp\big(ip\, \lambda_{m,\ell}^{(j)}\big)\ 
\sum_{\beta=1}^{\cL}\ \Big(\wh \Psi_{-1}(p)\Big)^{-1}_{\alpha\beta}
\wh\Psi_{-1}^{(n_\beta,m)}(p)\;.
\end{equation}
From the lemma \ref{lem:inv}, the sum on $\beta$ in the previous equation 
can be computed and we get finally
\begin{equation}
\wh\fc_{n_\alpha}^{(j)}(p)=-\frac{1}{2\pi}\sump_{m=n_{\alpha-1}+\half}^{n_\alpha-\half}
\sum_{\ell=1}^{\wt \nu_m^{(j)}}
\wh\omega^-_{m}(p)\,\exp(ip\lambda_{m,\ell}^{(j)})
-\frac{1}{2\pi}\sump_{m=n_{\alpha}+\half}^{n_{\alpha+1}-\half}
\sum_{\ell=1}^{\wt \nu_m^{(j)}}
\wh\omega^+_{m}(p)\,\exp(ip\lambda_{m,\ell}^{(j)})\;,
\end{equation}
The final result is obtained through inverse Fourier transform of this 
last equality.
\finprf
In theorem \ref{th:cj}, corrections to densities have been expressed 
in terms of the parameters $\lambda_{m,\ell}^{(j)}$ ($m\in \overline \cN$). 
These parameters are implicit functions of 
$\wt\lambda_{n,\ell}^{(j)}$, as stated in the 
following theorem.
\begin{theorem}\label{th:constraint}
The relations between the Bethe roots corresponding to holes in the filled 
seas and the ones corresponding to new strings are given by, for $m\in\overline\cN$,
\begin{eqnarray}
\label{eq:constraint}
 2\pi\,Q_{m,k}^{(j)}&=&
\sum_{d=1}^{\cD^{(j)}_{\gamma^-(m)}}
\Omega^{+}_{m}(\lambda_{m,k}^{(j)}-\wt\lambda_{\gamma^-(m),d}^{(j)})
+\sum_{d=1}^{\cD^{(j)}_{\gamma^+(m)}}
\Omega^-_{m}(\lambda_{m,k}^{(j)}-\wt\lambda_{\gamma^+(m),d}^{(j)})
\\
&&
\hspace{-2cm}-\sump_{r=\gamma^-(m)+\half}^{\gamma^+(m)-\half}\left\{
\sum_{\ell=1}^{\wt\nu^{(j-1)}_{r}}
F^{(r,m)}_{-1}(\lambda_{m,k}^{(j)}-\lambda_{r,\ell}^{(j-1)})
+\sum_{\ell=1}^{\wt\nu^{(j)}_{r}}
F^{(r,m)}_{2}(\lambda_{m,k}^{(j)}-\lambda_{r,\ell}^{(j)})
+\sum_{\ell=1}^{\wt\nu^{(j+1)}_{r}}
F^{(r,m)}_{-1}(\lambda_{m,k}^{(j)}-\lambda_{r,\ell}^{(j+1)})
\right\}\nonumber
\end{eqnarray}
where the functions $\Omega^\pm_m$ are a primitive of $\omega^\pm_m$
\begin{equation}
\Omega^\pm_m(\lambda)=2\arctan\Big[
\tan\Big(\frac{\pi}{2}\;\frac{\vert m-\gamma^\pm(m)\vert}{\gamma^+(m)-\gamma^-(m)}\Big)
\tanh\Big(\frac{\pi}{2}\;\frac{\lambda}{\gamma^+(m)-\gamma^-(m)}\Big)
\Big]
\end{equation}
and the functions $F_a^{(r,m)}(\lambda)$ are
\begin{eqnarray}
\label{eq:F1}
F_{-1}^{(r,m)}(\lambda)&=&
-\sum_{q=\vert r-m\vert+\half}^{r+m-2\gamma^-(m)-\half}
\Gamma^{(r)}_{q}(\lambda)\\
\label{eq:F2}
F_{2}^{(r,m)}(\lambda)&=&
\begin{cases} \displaystyle
\Gamma^{(r)}_{2m-2\gamma^-(m)}(\lambda)\,
+2\,\sum_{q=1}^{2m-1-2\gamma^-(m)} \Gamma^{(r)}_{q}(\lambda)\,
 &\mbox{ if }m= r\\[1.2ex]
\displaystyle \Gamma^{(r)}_{r+m-2\gamma^-(m)}(\lambda)\,+
\,\Gamma^{(r)}_{\vert r-m\vert}(\lambda)\,
+2\sum_{q=\vert r-m\vert+1}^{r+m-1-2\gamma^-(m)}
\Gamma^{(r)}_{q}(\lambda)
&\mbox{ if }m\neq r\end{cases}
\\
\Gamma^{(r)}_{q}(\lambda) &=& -2\arctan\left[
\tanh\Big(\frac{\pi}{2}\;\frac{\lambda}{\gamma^+(r)-\gamma^-(r)}\Big)
\tan\Big(\frac{\pi}{2}\;\frac{q-\gamma^+(r)+\gamma^-(r)}{\gamma^+(r)-\gamma^-(r)}\Big)\right]
\end{eqnarray}
\end{theorem}
\prf
Postponed in appendix \ref{ap:proof}.
\finprf
Let us remark that if $m,r\in\overline\cN$ are such that $m,r>n_\cL$, the functions 
$\Omega^+_m(\lambda)$, $\Gamma^{(m)}_{q}(\lambda)$ and $F_a^{(r,m)}(\lambda)$ ($a=-1,2$) 
reduce to 
\begin{equation}
\Omega^+_m(\lambda)=2\arctan\Big(\frac{\lambda}{m-n_\cL}\Big)\mb{,}
\Gamma^{(m)}_{q}(\lambda) =2\arctan\Big(\frac{\lambda}{q}\Big)=\varphi_{2q}(\lambda)
\end{equation} 
and
\begin{equation}
F_a^{(r,m)}(\lambda)=\Phi_a^{(r-n_\cL-\half,m-n_\cL-\half)}(\lambda)\;.
\end{equation}
Notice also that $F^{(r,m)}(\lambda)=F^{(m,r)}(\lambda)$ when 
$\gamma^-(m)=\gamma^-(r)$,
which is always the case in the relations where these functions are used.

\subsection{Examples}

\textit{1.} For $gl(N)$ fundamental spin chain, we get 
$\overline \cN=\{\half,1,\frac{3}{2},\dots\}$ and $\gamma^-(m)=0$,  
$\gamma^+(m)=\infty$ for any $m\in\overline \cN$. Thus, the corrections 
for the real Bethe roots reduce to
\begin{equation}
\fc_{0}^{(j)}(\lambda)=
-\frac{1}{\pi}\ \sump_{m=\half}^{\infty}
\sum_{\ell=1}^{\wt \nu_m^{(j)}}
\frac{m}{(\lambda-\lambda_{m,\ell}^{(j)})^2+m^2}
\end{equation}
The constraints (\ref{eq:constraint}) become, for $m\in\overline\cN$,
\begin{eqnarray}
2\pi Q_{m,k}^{(j)}&=&\sum_{d=1}^{\cD_0^{(j)}}
\varphi_{2m}(\lambda_{m,k}^{(j)}-\wt\lambda_{d}^{(j)})-\sump_{r=\half}^\infty\left\{
\sum_{\ell=1}^{\wt\nu_r^{(j-1)}}
\Phi_{-1}^{(r-\half,m-\half)}(\lambda_{m,k}^{(j)}-\lambda_{r,\ell}^{(j-1)})
\right.\\
&&\hspace{2cm}\left.
+\sum_{\ell=1}^{\wt\nu_r^{(j)}}
\Phi_{2}^{(r-\half,m-\half)}(\lambda_{m,k}^{(j)}-\lambda_{r,\ell}^{(j)})
+\sum_{\ell=1}^{\wt\nu_r^{(j+1)}}
\Phi_{-1}^{(r-\half,m-\half)}(\lambda_{m,k}^{(j)}-\lambda_{r,\ell}^{(j+1)})
\right\}\nonumber
\end{eqnarray}
We remind that $\wt\lambda_{d}^{(j)}$ stand for $\wt\lambda_{0,d}^{(j)}$.

For the particular case $\wt \nu^{(j)}_m=0$ for any $m\geq 1$ (i.e. only 
strings of length 2 are added), we get
\begin{eqnarray}
2\pi Q_{\half,k}^{(j)}&=& \sum_{\ell=1}^{\wt\nu_\half^{(j)}}
\varphi_{1}(\lambda_{\half,k}^{(j)}-\lambda_{\half,\ell}^{(j-1)})
-\sum_{\ell=1}^{\wt\nu_\half^{(j)}}
\varphi_{2}(\lambda_{\half,k}^{(j)}-\lambda_{\half,\ell}^{(j)})
+\sum_{\ell=1}^{\wt\nu_\half^{(j)}}
\varphi_{1}(\lambda_{\half,k}^{(j)}-\lambda_{\half,\ell}^{(j+1)})
\nonumber\\
&& +
\sum_{d=1}^{\cD_0^{(j)}}
\varphi_{1}(\lambda_{\half,k}^{(j)}-\wt\lambda_{d}^{(j)})
\end{eqnarray}
We recover a previous result computed in \cite{anne}.\\

\textit{2.} For a $gl(2)$ spin $s$ spin chain, we get
$\overline \cN=\{0,\half,\dots,s-\frac{3}{2},s-1,s,s+\half,\dots\}$.
Thus $\gamma^+(m)=s-\half$, $\gamma^-(m)=-\half$ for $0\leq m\leq s-1$ and 
$\gamma^+(m)=\infty$, $\gamma^-(m)=s-\half$ for $s\leq m$. The
corrections are
\begin{equation}
\fc_{s-\half}(\lambda)=-\sump_{m=0}^{s-1}
\sum_{\ell=1}^{\wt \nu_m}
\frac{1}{2s}\;\frac{\sin(\pi\frac{2m+1}{2s})}
{\cosh(\pi\frac{\lambda-\lambda_{m,\ell}}{s})
+\cos(\pi\frac{2m+1}{2s})}
-\sump_{m=s}^{\infty}
\sum_{\ell=1}^{\wt \nu_m}
\frac{1}{\pi}\frac{m-s+\half}
{(\lambda-\lambda_{m,\ell})^2+(m-s+\half)^2}
\end{equation}
The constraints become, for $m=0,\half,\dots,s-1$,
\begin{equation}
2\pi Q_{m,k}=2\sum_{d=1}^{\cD_{s-\half}}
\arctan\left[\tan\left(\frac{\pi m}{2s-1}\right)
\tanh\left(\frac{\pi(\lambda_{m,k}-\wt\lambda_{d})}
{2s-1}\right)\right]-\sump_{r=\half}^{s-1} \sum_{\atopn{\ell=1}
{\ell\neq k}}^{\wt\nu_r}F_2^{(r,m)}(\lambda_{m,k}-\lambda_{r,\ell})
\end{equation}
and, for $m=s,s+\half,\dots$,
\begin{equation}
2\pi Q_{m,k}=2\sum_{d=1}^{\cD_{s-\half}}
\arctan\left(\frac{\lambda_{m,k}-\wt\lambda_{d}}
{m-s+\half}\right)
-\sump_{r=s}^\infty \sum_{\ell=1}^{\wt\nu_r}
\Phi_2^{(r-s,m-s)}(\lambda_{m,k}-\lambda_{r,\ell})
\end{equation}
keeping the notation 
$\wt\lambda_{d}\equiv\wt\lambda_{s-\half,d}$ for the Bethe roots of 
the holes.

\textit{3.} For an alternating spin chain with $n_1<n_2$, we get
$\gamma^-(m)=-\half$, $\gamma^+(m)=n_1$ for $0\leq m \leq n_1-\half$, 
$\gamma^-(m)=n_1$, $\gamma^+(m)=n_2$ for $n_1+\half\leq m \leq n_2-\half$ and
$\gamma^-(m)=n_2$, $\gamma^+(m)=\infty$ for $n_2+\half\leq m$. Thus,
the corrections to the densities are
\begin{eqnarray}
\fc_{n_1}^{(j)}(\lambda)
&=&-\half\sump_{m=0}^{n_1-\half}\sum_{\ell=1}^{\wt \nu_m^{(j)}}
\frac{1}{n_1+\half}\;\frac{\sin\left(\pi\frac{m+n_1+\half}{n_1+\half}\right)}
{\cosh\left(\pi\frac{\lambda-\lambda_{m,\ell}^{(j)}}{n_1+\half}\right)
+\cos\left(\pi\frac{m+n_1+\half}{n_1+\half}\right)}\\
&&-\half\sump_{m=n_1+\half}^{n_2-\half}\sum_{\ell=1}^{\wt \nu_m^{(j)}}
\frac{1}{n_2-n_1}\;\frac{\sin\left(\pi\frac{m+n_2-n_1}{n_2-n_1}\right)}
{\cosh\left(\pi\frac{\lambda-\lambda_{m,\ell}^{(j)}}{n_2-n_1}\right)
+\cos\left(\pi\frac{m+n_2-n_1}{n_2-n_1}\right)}\nonumber
\end{eqnarray}
and
\begin{eqnarray}
\fc_{n_2}^{(j)}(\lambda) &=&
-\frac{1}{2(n_2-n_1)}\sump_{m=n_1+\half}^{n_2-\half}\sum_{\ell=1}^{\wt \nu_m^{(j)}}
\;\frac{\sin\left(\pi\frac{m+n_2-n_1}{n_2-n_1}\right)}
{\cosh\left(\pi\frac{\lambda-\lambda_{m,\ell}^{(j)}}{n_2-n_1}\right)
+\cos\left(\pi\frac{m+n_2-n_1}{n_2-n_1}\right)}\\
&&-\frac{1}{\pi}\sump_{m=n_2+\half}^{\infty}\sum_{\ell=1}^{\wt \nu_m^{(j)}}
\frac{m-n_2}
{(\lambda-\lambda_{m,\ell}^{(j)})^2+(m-n_2)^2}\nonumber
\end{eqnarray}
The constraints between the roots are still given by the theorem 
\ref{th:constraint}. Since it is a rather cumbersome equation,
we refrain from giving its explicit form.

\subsection{Excitations within the vacuum sector \label{sec:spin-min}}
We consider an excitation in the vacuum sector, as defined in remark 
\ref{rmk2}: its  hole configuration is given by
$\{\wt\mu_{m}^{(j)}\,,\ m\in\cN\,,\ 
j=1,\ldots,N-1\}$, where $\wt\mu_{m}^{(j)}=N\,\mu_{m}^{(j)}$ 
and $\{\mu_{m}^{(j)}\}$ describes the hole configuration
 as given in lemma \ref{lem1}. To this 
configuration, if we suppose that $n_{1}>0$, one can add new strings 
of the following type. 

We first introduce (for $m\in\bcN$):
\begin{equation}
\wt w_{m}^{(j)}=\sum_{\alpha=1}^\cL \delta_{m,\bar 
n_{\alpha}}\,v_{n_{\alpha}}^{(j)}
\mb{where}
\bar n_{\alpha}=\max(m\in\bcN\,,\ m<n_{\alpha})\,. 
\end{equation}
We remind that 
$v_{m}^{(j)}=2\,\wt\mu_{m}^{(j)}-\wt\mu_{m}^{(j-1)}-\wt\mu_{m}^{(j+1)}$ 
($m\in\bcN$) and 
$\wt w_{n}^{(j)}=2\,\wt \nu_{n}^{(j)}-\wt \nu_{n}^{(j-1)}
-\wt \nu_{n}^{(j+1)}$.

The set $\{\mu_{m}^{(j)}\,,\ \wt \nu_{n}^{(j)}\,,\ m\in\cN\,,\ n\in\bcN\,,\ 
j=1,\ldots,N-1\}$ define a configuration with a spin
\begin{eqnarray}
S_{j} &=& \sum_{\alpha=1}^\cL 2(n_{\alpha}-\bar n_{\alpha})\,v_{n_{\alpha}}^{(j)}
\end{eqnarray}
and valences
\begin{equation}\begin{array}{lcll}
\displaystyle
\wt P_{m}^{(j)} &=& \displaystyle\wt\nu_{m}^{(j)} &\mb{if} 
m<n_{1}\\[1.2ex]
\displaystyle
\wt P_{m}^{(j)} &=& \displaystyle\wt\nu_{m}^{(j)} +  \sum_{\alpha=\beta_{1}+1}^{\beta_{2}} 
2\,(m-\bar n_{\alpha})\,v_{n_{\alpha}}^{(j)} +  \sum_{\alpha=\beta_{2}+1}^{\cL} 
2\,(n_{\alpha}-\bar n_{\alpha})\,v_{n_{\alpha}}^{(j)}
&\mb{if} n_{\beta_{1}}\leq m\leq\bar n_{\beta_{2}}\\[1.2ex]
\displaystyle
\wt P_{m}^{(j)} &=& \displaystyle\wt\nu_{m}^{(j)} +
\sum_{\alpha=1}^\cL 2(n_{\alpha}-\bar n_{\alpha})\,v_{n_{\alpha}}^{(j)}
&\mb{if} n_{\cL}\leq m
\end{array}
\end{equation}
Above, $\beta_{1}$ and $\beta_{2}$ depend on $m$ and are such that
\begin{equation}
 n_{\beta_{1}}\leq m< n_{\beta_{1}+1}\mb{and} 
\bar n_{\beta_{2}-1}< m\leq\bar n_{\beta_{2}}
\end{equation}
This construction generalizes the construction given in \cite{Tak} 
for the spin $s$ XXX spin chain.

\subsubsection*{Example} For the $gl(2)$ spin $s$ spin chain, the lemma 
\ref{lem1} 
hole excitation is given by 
$\wt\nu_{s-\half}=\frac{L-2}{2}$, leading to $\wt 
w_{n}=2\delta_{s-1,n}$, i.e. $\wt\nu_{s-1}=1$. 
Then we get an excited state with spin $\frac{S}{2}=1$, 
corresponding to the adjoint representation of $sl(2)$. As already 
mentioned, we recover the results given in \cite{Tak}. 

\section{Hamiltonian and energies\label{sec:energies}}
\subsection{$L_{0}$-local Hamiltonians}
\begin{definition}
An Hamiltonian is called $L_{0}$-local when it possesses only interactions 
between neighbours separated by at most 
$L_{0}-1$ sites. 
\end{definition}
This definition is quite natural. It is used in the following: 
\begin{conjecture}
For any  $L_{0}$-regular spin chain there always exists an Hamiltonian 
which is $L_{0}$-local.
\end{conjecture}
This conjecture is supported by the theorem \ref{theo:L0-loc} given 
below. To 
formulate it, we first introduce local representations of $gl(N)$.
On each site $\ell$ we denote the representation of the $gl(N)$ 
algebra by 
\begin{equation}
\cE_{1,\ell}=\sum_{i,j=1}^N
E_{ij}\otimes e^{(\ell)}_{ji}
\end{equation}
 where the indices label the auxiliary 
and/or quantum spaces, the generator acting as identity operator in 
the other spaces. Accordingly, the 
Yangian representation at site $\ell$ will be 
$T_{1,\ell}(\lambda)=\lambda\,\II_{N}\otimes \II
+i\cE_{1,\ell}$.

\begin{theorem}\label{theo:L0-loc}
For a $gl(N)$ spin chain, $L_{0}$-regular, with a fundamental
representation at sites $1+q\,L_{0}$, $q\in\ZZ_{+}$, the Hamiltonian 
$H=-i\,t(0)^{-1}\,t'(0)$ is $L_{0}$-local and takes the form:
\begin{eqnarray}
H_{<1,2,\ldots, L>} &=& \sum_{q=1}^p\left\{
\left(\prod_{\ell=2}^{\atopn{\longrightarrow}{L_0}}
\cE_{1+qL_{0},\ell+(q-1)L_{0}}\right)^{-1}\ 
\left(\prod_{\ell=2}^{\atopn{\longrightarrow}{L_0}}
\cE_{1+(q-1)L_{0},\ell+(q-1)L_{0}}\right)\ P_{1+(q-1)L_{0},1+qL_{0}} 
\right.\nonu
&& \qquad -\left.\sum_{\ell_{1}=2}^{L_{0}}\left(\prod_{\ell=2}^{\atopn{\longrightarrow}{L_0}}
\cE_{1+qL_{0},\ell}\right)^{-1}\ 
\left(\prod_{\atopn{\ell=2}{\ell\neq \ell_{1}}}^{\atopn{\longrightarrow}{L_0}}
\cE_{1+qL_{0},\ell}\right)\right\}
\label{eq:Ham}
\end{eqnarray}
where the total number of sites is $L=pL_{0}$.
\end{theorem}
\prf
We consider a $gl(N)$ spin chain, $L_{0}$-regular, with a fundamental
representation at sites $1+q\,L_{0}$, $q\in\ZZ_{+}$. We define the
Hamiltonian as
\begin{eqnarray}
H=-i\,t(0)^{-1}\,t'(0) \mb{with} t'(\lambda)=\frac{d}{d\lambda}t(\lambda)
\end{eqnarray}
Starting with $t(\lambda)=tr_{0}\left(T_{01}(\lambda)\ldots 
T_{0L}(\lambda)\right)$, we decompose $t'(0)$ into three terms, 
depending whether the 
derivation applies on $T_{01}(\lambda)$, on $T_{0,\ell}(\lambda)$, 
$1<\ell<L_{0}+1$ or on $T_{0,\ell}(\lambda)$, 
$L_{0}<\ell$. After multiplication by $t(0)^{-1}$, the first two
terms can be easily computed, while in the last one, we recognize the 
Hamiltonian of the same type of spin chain, but where the 
first $L_{0}$ sites have been suppressed.
Then, we get the recursion formula for the Hamiltonian:
\begin{eqnarray}
H_{<1,2,\ldots, L>} &=& H_{<L_{0}+1,L_{0}+2,\ldots, L>}
+\left(\prod_{\ell=2}^{\atopn{\longrightarrow}{L_0}}
\cE_{1+L_{0},\ell}\right)^{-1}\ 
\left(\prod_{\ell=2}^{\atopn{\longrightarrow}{L_0}}
\cE_{1,\ell}\right)\ P_{1,1+L_{0}} \nonu
&&-\sum_{\ell_{1}=2}^{L_{0}}\left(\prod_{\ell=2}^{\atopn{\longrightarrow}{L_0}}
\cE_{1+L_{0},\ell}\right)^{-1}\ 
\left(\prod_{\atopn{\ell=2}{\ell\neq \ell_{1}}}^{\atopn{\longrightarrow}{L_0}}
\cE_{1+L_{0},\ell}\right)
\end{eqnarray}
This recursion is easily solved, and we get (\ref{eq:Ham}).\finprf
Of course, depending on the representations sitting at sites 
$\ell\neq 1+qL_{0}$, further simplifications may arise in the 
expression (\ref{eq:Ham}), see examples below.

Note that once the representations of the spin chain are fixed, 
$\cE_{1,\ell}$ becomes a matrix, which can be easily inverted.
In a more algebraic way, one 
 may compute $\left(\cE_{1,\ell}\right)^{-1}$ using the
quantum comatrix in the Yangian $Y(N)$. 

When no fundamental representation occurs on the chain, one has to 
deal with a fusion procedure, in order to get auxiliary space with 
higher dimension (see e.g. \cite{KS,MTV}). 
It is reasonable to think that to get a $L_{0}$-local Hamiltonian, the  
auxiliary space has to be isomorphic to one of the representation on the 
chain. Then, the transfer matrix will be of different type.
However,  the Bethe 
equations (and thus the results presented above) will remain 
unchanged, since 
all fused transfer matrices commute among themselves.

\subsubsection*{Examples}

\quad\textit{1.} For $gl(N)$ fundamental spin chain, one recovers the usual spin 
chain Hamiltonian
\begin{eqnarray}
H=\sum_{\ell=1}^L P_{\ell,\ell+1}
\end{eqnarray}

\textit{2.} The spin-$s$ $gl(2)$ chain does not enter the hypothesis. 
However, the existence of a local (1-local) Hamiltonian is known 
\cite{TTF,fafa}. 

\textit{3.} For an alternating spin chain (with $L=2p$), with fundamental 
representations on odd sites, one gets an 
Hamiltonian with next-to-nearest neighbours interactions (i.e. 
2-local in our terminology):
\begin{eqnarray}
H=\sum_{\ell=1}^{L/2}\left\{ \left(\cE_{2\ell+1,2\ell}\right)^{-1}\, 
\cE_{2\ell-1,2\ell}\, P_{2\ell-1,2\ell+1}-\cE_{2\ell+1,2\ell}
\right\}
\end{eqnarray}

If furthermore, one takes a $gl(2)$ spin chain, the matrices $\cE$ 
and $\cE^{-1}$ are easy to compute.
 Indeed, in a $sl(2)$ representation of spin $s$ (square matrices of size 
$2s+1$), one has 
\begin{eqnarray}
\pi_{s}(e_3) = \sum_{n=1}^{2s+1}\big(s+1-n\big)\,E^{(s)}_{nn}
\mb{;}
\pi_{s}(e_{+}) = \sum_{n=1}^{2s}\sqrt{n(2s+1-n)}
\,E^{(s)}_{n,n+1}
\nonu
\pi_{s}(e_{-}) = \sum_{n=1}^{2s} \sqrt{n(2s+1-n)}
\,E^{(s)}_{n+1,n}
\mb{;}
\pi_{s}(1) = \sum_{n=1}^{2s+1} E^{(s)}_{nn} = \II_{2s+1}
\end{eqnarray}
where $E^{(s)}_{ij}$ are the $(2s+1)\times(2s+1)$ elementary matrices with 
1 at position $(i,j)$ and 0 elsewhere. In this representation, one has
\begin{eqnarray}
[e_3,e_{\pm}]=\pm\,e_{\pm}\mb{;} [e_{+},e_{-}]=2\,e_3 
\mb{and}tr(e_{+}\,e_{-})=tr(e_3^2)=\frac{s(s+1)(2s+1)}{3}
\end{eqnarray}
Hence, we get
\begin{eqnarray}
2\,\cE_{2\ell+1,2\ell} = \II_{2}\otimes \II_{2s+1}+ 
4\,\Big(\sigma_3\otimes \pi_{s}(e_3) +
 \half\big(\sigma_{+}\otimes \pi_{s}(e_{-}) 
+\sigma_{-}\otimes \pi_{s}(e_{+}) \big)
\Big) 
\equiv \II+4\,X_{s}
\end{eqnarray}
where we have introduced the Pauli matrices
\begin{eqnarray}
\sigma_3= \left(\begin{array}{cc} 
\half & 0 \\ 0 & -\half 
\end{array}\right)\mb{;}
\sigma_{+}= \left(\begin{array}{cc} 
0 & 1 \\ 0 & 0 
\end{array}\right)\mb{;}
\sigma_{-}= \left(\begin{array}{cc} 
0 & 0 \\ 1  & 0 
\end{array}\right)\,. 
\end{eqnarray}
Due to the properties of Pauli matrices, we have (whatever 
the value of $s$)
\begin{eqnarray}
X_{s}^2 = \frac{1}{4}\II_{2}\otimes C_{s}-\half\,X_{s}
\end{eqnarray}
where $C_{s}$ is the value of the Casimir operator in the spin $s$ 
representation:
\begin{eqnarray}
C_{s}=\pi_{s}\Big(e_3^2 +\half\big(e_{-}e_{+}+e_{+}e_{-}\big)\Big)
= s(s+1)\, \II_{2s+1}\,.
\end{eqnarray}
This implies that 
\begin{eqnarray}
(\cE_{2\ell+1,2\ell})^{2}  = \Big(s+\half \Big)^2\,
\II_{2}\otimes \II_{2s+1}
\quad\Rightarrow\quad 
(\cE_{2\ell+1,2\ell})^{-1}  = 
\frac{1}{\big(s+\half \big)^2}\,\cE_{2\ell+1,2\ell}
\end{eqnarray}
Using the notation
\begin{eqnarray}
\vec{\sigma}_{2\ell-1}\cdot \vec{S}_{2\ell}
=\sigma^{(2\ell-1)}_3\otimes \pi_{s}^{(2\ell)}(e_3) +
 \sigma^{(2\ell-1)}_{+}\otimes \pi_{s}^{(2\ell)}(e_{-}) 
+\sigma^{(2\ell-1)}_{-}\otimes \pi_{s}^{(2\ell)}(e_{+}) 
\end{eqnarray}
where $\sigma_{3,\pm}^{(2\ell-1)}$ and $\pi_{s}^{(2\ell)}(e_{3,\pm})$
are representation of $e_{3,\pm}$ at site $2\ell-1$ (spin $\half$) 
and $2\ell$ (spin $s$) resp., 
and obvious generalisations of this notation to other cases, we can rewrite the 
Hamiltonian as
\begin{eqnarray}
H &=& \frac{-1}{2\big(s+\half\big)^2}\,\sum_{\ell=1}^{L/2}\Big\{ 
s(s+1) +4s(s+1)\,\vec{S}_{2\ell}\cdot \vec{\sigma}_{2\ell+1}-
\vec{\sigma}_{2\ell-1}\cdot \vec{S}_{2\ell} -
\vec{\sigma}_{2\ell-1}\cdot \vec{\sigma}_{2\ell+1}\nonu
&&-4\,(\vec{S}_{2\ell}\cdot \vec{\sigma}_{2\ell+1}) 
(\vec{\sigma}_{2\ell-1}\cdot \vec{S}_{2\ell})
-4\,(\vec{S}_{2\ell}\cdot \vec{\sigma}_{2\ell+1}) 
(\vec{\sigma}_{2\ell-1}\cdot \vec{\sigma}_{2\ell+1})
-4\,(\vec{\sigma}_{2\ell-1}\cdot \vec{S}_{2\ell})
(\vec{\sigma}_{2\ell-1}\cdot \vec{\sigma}_{2\ell+1})
\nonu
&&-16\,(\vec{S}_{2\ell}\cdot \vec{\sigma}_{2\ell+1}) 
(\vec{\sigma}_{2\ell-1}\cdot \vec{S}_{2\ell})
(\vec{\sigma}_{2\ell-1}\cdot \vec{\sigma}_{2\ell+1})
\Big\}
\end{eqnarray}
which describes an alternating $gl(2)$ spin chains with spins $\half$ 
and $s$ and next-to-nearest neighbours interaction.

\subsection{Energies\label{sec:energy}}
Motivated by the result given above, we focus here on the case of 
$L_0$-regular spin chain containing at least one
fundamental representation. This implies that $n_{1}=0$ (we remind 
that the $n_i$'s are ordered). 
In this case, the energy 
(eigenvalue of the Hamiltonian (\ref{eq:Ham})) 
per site for a state characterized by the densities
$\ft_{n}^{(j)}(\lambda)$, $j=1,\ldots N-1$, $n\in\half\ZZ_{+}$ (as 
given in (\ref{eq:correction}) and (\ref{eq:correction2}))
reads
\begin{equation}
E-E_{+}=\sump_{m\in \ZZ_{+}/2}\ \int_{-\infty}^{+\infty} d\lambda\
\ft_{m}^{(1)}(\lambda)\ \frac{2m+1}{\lambda^{2}+(\frac{2m+1}{2})^{2}}
\end{equation}
where $E_{+}$ is the energy of the pseudo-vacuum
\begin{equation}
E_{+}=-\frac{2}{L_{0}}\sum_{\alpha=1}^\cL \sum_{\ell\in I_{\alpha}}
\frac{1}{\bar a_{\alpha}+j_{\ell}}
\end{equation}
In particular, for the vacuum state (i.e. when densities 
 are given by (\ref{eq:densitemagique})), one gets \cite{thermy}
\begin{equation}
E-E_{+}=\frac{2}{NL_{0}}\sum_{\alpha=1}^\cL \sum_{\ell\in I_{\alpha}}\left\{
\psi(\frac{\bar a_{\alpha}-j_{\ell}+1}{2N}+1)
-\psi(\frac{\bar a_{\alpha}+j_{\ell}}{2N})\right\}
\end{equation}
where $\psi(x)=\frac{d}{dx}\ln\Gamma(x)$ is the Euler digamma 
function. 
For the excited states studied in this paper, the correction to this energy are 
given by the following theorem.
\begin{theorem}
For a $L_0$-regular spin chain with, at least, one fundamental 
representation, the correction at order $1/L$ to the energy of the 
vacuum, due to the general excited states introduced in section \ref{sec:gh}, is
\begin{equation}
\Delta E= \frac{2\pi}{N}\, \sum_{k=1}^{N-1}\ \sum_{d=1}^{\cD_{0}^{(k)}}\
   \frac{\sin\left(\frac{k\pi}{N}\right)}
      {\cos(\frac{k\pi}{N})-\cosh(\frac{2 \pi}{N}\wt\lambda_{0,d}^{(k)})}
\label{eq:DEh}
\end{equation}
\end{theorem}
\prf
The corrections at order $1/L$ to the energy, 
due to the modifications of the densities in the seas $m\in\cN$, are, respectively,
\begin{equation}
\Delta^h E=\sump_{m\in \cN}\ \int_{-\infty}^{+\infty} d\lambda\
\rho_{m}^{(1)}(\lambda)\ \frac{2m+1}{\lambda^{2}+(\frac{2m+1}{2})^{2}}
\mb{and}
\Delta^s E=\sump_{m\in \cN}\ \int_{-\infty}^{+\infty} d\lambda\
\fc_{m}^{(1)}(\lambda)\ \frac{2m+1}{\lambda^{2}+(\frac{2m+1}{2})^{2}}
\end{equation}
which can be computed via the Plancherel's theorem
\begin{equation}
\Delta^h E=2\pi\sump_{m\in \cN}\ \int_{-\infty}^{+\infty} dp\
\wh \rho_{m}^{(1)}(p)\ e^{-\frac{2m+1}{2}\vert p \vert}
\mb{and}
\Delta^s E=2\pi\sump_{m\in \cN}\ \int_{-\infty}^{+\infty} dp\
\wh\fc_{m}^{(1)}(p)\ e^{-\frac{2m+1}{2}\vert p \vert}\; .
\end{equation}
The corrections to the energy 
provided by the densities (\ref{eq:correction2}) of new strings read
\begin{equation}
\Delta^n E=\sump_{m\in\bcN}\ \sum_{\ell=1}^{\tilde \nu_m^{(1)}}\
\frac{2m+1}{(\lambda_{m,\ell}^{(1)})^2+(\frac{2m+1}{2})^2}\;.
\end{equation}
Using the explicit forms (\ref{eq:densite-correction}) of the functions 
$\wh\rho_{m}^{(1)}(p)$, we can show 
that, in the expression of $\Delta^h E$, the coefficent of 
$\exp(i p\,\wt\lambda_{n_{\alpha},d}^{(k)})$ identically vanishes for 
all $n_\alpha$ but $n_1$. Thus, one
obtains (reminding that $n_{1}=0$):
\begin{eqnarray}
\label{eq:Dhtf}
\Delta^h E &=&  -\sum_{k=1}^{N-1}\sum_{d=1}^{\cD_{0}^{(k)}} 
\int_{-\infty}^{\infty} dp \ 
\frac{\sinh((N-k)\frac{\vert p\vert}{2})}
{\sinh(\frac{N\vert p\vert}{2}) }
\;e^{i p\,\wt\lambda_{0,d}^{(k)}}\;.
\end{eqnarray}
Similarly, using the explicit forms (\ref{eq:c}) of 
$\wh\fc_{m}^{(1)}(p)$,  we 
can simplify $\Delta^s E$:
\begin{eqnarray}
\label{eq:Dstf}
\Delta^s E &=& 
-\sump_{m\in\bcN}\sum_{\ell=1}^{\tilde \nu_m^{(1)}}
\int_{-\infty}^{\infty} dp \ 
e^{-\frac{2m+1}{2}\vert p \vert}
e^{ip\lambda_{m,\ell}^{(1)}}\qquad
\end{eqnarray}
Then, performing an inverse Fourier transform of (\ref{eq:Dhtf}) and (\ref{eq:Dstf}),
we get the results (\ref{eq:DEh}) by remarking that $\Delta^s E=-\Delta^n E$.
\finprf

We would like to emphasize that although the studied spin chains 
are quite general, the final expression (\ref{eq:DEh})
for the corrections $\Delta E$ to the energy are very simple.
Indeed, these corrections have the same shape for any $\tilde
\lambda_{0,d}^{(k)}$,  and  can rewritten as:
\begin{equation}
\Delta E=\sum_{k=1}^{N-1} \sum_{d=1}^{\cD_0^{(k)}}
\delta^{(k)}(\wt \lambda_{0,d}^{(k)})
\qmbox{where}\delta^{(k)}(\lambda)=\frac{2\pi}{N}\ \frac{\sin\left(\frac{k\pi}{N}\right)}
      {\cos(\frac{k\pi}{N})-\cosh(\frac{2 \pi}{N}\lambda)}\;.
\end{equation}
Therefore, the holes in the filled seas of real strings
can be interpreted as particle-like excitations with  energy
$\delta^{(k)}(\wt \lambda_{0,d}^{(k)})$ and rapidity $\wt \lambda_{0,d}^{(k)}$.

Moreover, since only real holes appear in the expression of
$\Delta E$, one can say that, at order $\frac{1}{L}$, the holes created in filled 
seas of non-real strings have a zero energy.
\appendix

\section{Proof of theorem \ref{th:constraint}\label{ap:proof}}

The parameters $\lambda_{n,\ell}^{(j)}$ satisfy the 
following constraints which are provided by the Bethe equations:
\begin{eqnarray}
\label{eq:bethebarN} 
&& \sump_{n\in\cN}\left\{
\int_{-\infty}^{\infty} 
d\lambda \,\Big({\ft}_{n}^{(j-1)}(\lambda)+
\ft_{n}^{(j+1)}(\lambda)\Big)\,
{\Phi_{-1}^{(m,n)}}(\lambda^{(j)}_{m,k}-\lambda)
+\int_{-\infty}^{\infty} d\lambda \,\ft_{n}^{(j)}(\lambda)\, 
{\Phi_{2}^{(m,n)}}(\lambda^{(j)}_{m,k}-\lambda)
\right\}\ =\ \qquad\nonumber
\\
&&-\frac{1}{L}\sump_{q\in\overline\cN}\left\{
\sum_{\ell=1}^{\wt\nu^{(j-1)}_{q}}
\Phi_{-1}^{(q,m)}(\lambda^{(j)}_{m,k}-\lambda^{(j-1)}_{q,\ell})+
\sum_{\ell=1}^{\wt\nu^{(j)}_{q}}
\Phi_{2}^{(q,m)}(\lambda^{(j)}_{m,k}-\lambda^{(j)}_{q,\ell})+
\sum_{\ell=1}^{\wt\nu^{(j+1)}_{q}}
\Phi_{-1}^{(q,m)}(\lambda^{(j)}_{m,k}-\lambda^{(j+1)}_{q,\ell})
\right\}\nonu
&&+\frac{1}{L_{0}}
\sum_{\alpha=1}^{\cL}\Big(\sum_{\ell\in I_\alpha}\delta_{j,j_{l}}\Big)\,
\Phi_{\bar a_{\alpha}}^{(m)}(\lambda^{(j)}_{m,k}) -
\frac{2\pi}{L}\,Q_{m,k}^{(j)} \label{eqA1}
\end{eqnarray}
where $m\in\overline\cN$, 
$1\leq k\leq \wt\nu^{(j)}_m$ and $1\leq j \leq N-1$.
To simplify this relation, we replace the densities $\ft_{n}^{(l)}(\lambda)$ 
in the l.h.s. of (\ref{eqA1}) by their values obtained from 
(\ref{eq:s}) and (\ref{eq:t}):
\begin{equation}
\ft^{(l)}_{n}(\lambda)=\sigma^{(l)}_{n}(\lambda)+\frac{1}{L}\rho^{(l)}_{n}(\lambda)
+\frac{1}{L}\fc^{(l)}_{n}(\lambda)\;.
\end{equation}
The three terms of the above sum are reduced in the three following 
lemmas:
\begin{lemma}\label{lemmaA1}
For $m\in \overline \cN$ and $\lambda_0\in\RR$, we have the identity
\begin{eqnarray}
\label{eq:bethebarN0} 
&& \sump_{n\in\cN}\left\{
\int_{-\infty}^{\infty} 
d\lambda \,\Big({\sigma}_{n}^{(j-1)}(\lambda)+
\sigma_{n}^{(j+1)}(\lambda)\Big)\,
{\Phi_{-1}^{(m,n)}}(\lambda_{0}-\lambda)
+\int_{-\infty}^{\infty} d\lambda \,\sigma_{n}^{(j)}(\lambda)\, 
{\Phi_{2}^{(m,n)}}(\lambda_{0}-\lambda)
\right\} = \nonu
&&\frac{1}{L_{0}}
\sum_{\alpha=1}^{\cL}\Big(\sum_{\ell\in I_\alpha}\delta_{j,j_{l}}\Big)\,
\Phi_{\bar a_{\alpha}}^{(m)}(\lambda_{0})
\end{eqnarray}
\end{lemma}
\prf
We derivate w.r.t. $\lambda_0$ the l.h.s. of (\ref{eq:bethebarN0}), 
perform a Fourier 
transform and use the explicit form of the Fourier transform of 
the vacuum densities (\ref{eq:densitemagique}). 
Then, we remark that it is equal to the Fourier transform of the derivative of the r.h.s. of 
(\ref{eq:bethebarN0}) using, in particular,
\begin{eqnarray}
&& h_{j-1,k}(x)+h_{j+1,k}(x)-2\cosh(x)\,h_{j,k}(x)
\ =\ -\delta_{j,k}\sinh(Nx)\sinh(x)\nonumber\\
&&\mb{where} 
h_{j,k}(x)=\sinh\big(x\,(N-\max[j,k])\big)\,\sinh\big(x\,\min[j,k]\big)\;.
\end{eqnarray}
This proves the equality up to a constant, which is fixed by 
considering the value $\lambda_0=0$ in the equation, and
remarking that the densities are even functions while 
$\Phi_j^{(m,n)}(\lambda)$ is odd. 
\finprf
\begin{lemma}\label{lemmaA2}
For $m\in\overline\cN$ and for $\lambda_0\in \RR$, we have the 
equality
\begin{eqnarray}
\label{eq:bethebarN1} 
&& \sump_{n\in\cN}\left\{
\int_{-\infty}^{\infty} 
d\lambda \,\Big({\rho}_{n}^{(j-1)}(\lambda)+
\rho_{n}^{(j+1)}(\lambda)\Big)\,
{\Phi_{-1}^{(m,n)}}(\lambda_{0}-\lambda)
+\int_{-\infty}^{\infty} d\lambda \,\rho_{n}^{(j)}(\lambda)\, 
{\Phi_{2}^{(m,n)}}(\lambda_{0}-\lambda)
\right\}=\nonumber
\\
&&-\left(\sum_{d=1}^{\cD^{(j)}_{\gamma^-(m)}}
\Omega^+_m(\lambda_0-\wt\lambda_{\gamma^-(m),d}^{(j)})
+\sum_{d=1}^{\cD^{(j)}_{\gamma^+(m)}}
\Omega^-_m(\lambda_0-\wt\lambda_{\gamma^+(m),d}^{(j)})\right)
\end{eqnarray}
\end{lemma}
\prf
Performing a derivation w.r.t. $\lambda_0$ and a Fourier 
transform of the l.h.s. of (\ref{eq:bethebarN1}), we get, using 
expression (\ref{eq:densite-correction}), the equality 
(\ref{eq:bethebarN1}) up to a constant. Considering the limit 
$\lambda_{0}\to\infty$, one shows that the constant vanishes.
\finprf

\begin{lemma}\label{lemmaA3}
For $m\in \overline \cN$, and for $\lambda_0\in \RR$, we have
\begin{eqnarray}
\label{eq:bethebarN2} 
&&\hspace{-1cm} \sump_{n\in\cN}\left\{
\int_{-\infty}^{\infty} 
d\lambda \,\Big({\fc}_{n}^{(j-1)}(\lambda)+
\fc_{n}^{(j+1)}(\lambda)\Big)\,
{\Phi_{-1}^{(m,n)}}(\lambda_{0}-\lambda)
+\int_{-\infty}^{\infty} d\lambda \,\fc_{n}^{(j)}(\lambda)\, 
{\Phi_{2}^{(m,n)}}(\lambda_{0}-\lambda)
\right\}=
\\
&&-\sump_{q\in\overline\cN}\left\{
\sum_{\ell=1}^{\wt\nu^{(j-1)}_{q}}
\Phi_{-1}^{(q,m)}(\lambda_{0}-\lambda^{(j-1)}_{q,\ell})+
\sum_{\ell=1}^{\wt\nu^{(j)}_{q}}
\Phi_{2}^{(q,m)}(\lambda_{0}-\lambda^{(j)}_{q,\ell})+
\sum_{\ell=1}^{\wt\nu^{(j+1)}_{q}}
\Phi_{-1}^{(q,m)}(\lambda_{0}-\lambda^{(j+1)}_{q,\ell})
\right\}\nonumber\\
&&+\sump_{r=\gamma^-(m)+\half}^{\gamma^+(m)-\half}\left\{
\sum_{\ell=1}^{\wt\nu^{(j-1)}_{r}}
F^{(r,m)}_{-1}(\lambda_0-\lambda_{r,\ell}^{(j-1)})
+\sum_{\ell=1}^{\wt\nu^{(j)}_{r}}
F_{2}^{(r,m)}(\lambda_0-\lambda_{r,\ell}^{(j)})
+\sum_{\ell=1}^{\wt\nu^{(j+1)}_{r}}
F_{-1}^{(r,m)}(\lambda_0-\lambda_{r,\ell}^{(j+1)})
\right\}\nonumber
\end{eqnarray}
where $F_{-1}^{(r,m)}(\lambda)$ and $F_{2}^{(r,m)}(\lambda)$ 
are given by (\ref{eq:F1}) and (\ref{eq:F2}).
\end{lemma}
\prf
Performing a derivation w.r.t. $\lambda_0$ and a Fourier 
transform of the l.h.s. of (\ref{eq:bethebarN2}), we get, using 
expression (\ref{eq:c})

\begin{eqnarray}
&&-\sump_{q\in\overline\cN}\left\{
\sum_{\ell=1}^{\wt\nu^{(j-1)}_{q}}
\wh\Psi_{-1}^{(q,m)}(p)\,\exp(ip\lambda^{(j-1)}_{q,\ell})+
\sum_{\ell=1}^{\wt\nu^{(j)}_{q}}
\wh\Psi_{2}^{(q,m)}(p)\,\exp(ip\lambda^{(j)}_{q,\ell})+
\sum_{\ell=1}^{\wt\nu^{(j+1)}_{q}}
\wh\Psi_{-1}^{(q,m)}(p)\,\exp(ip\lambda^{(j+1)}_{q,\ell})
\right\}\nonumber\\
&&+\sump_{r=\gamma^-(m)+\half}^{\gamma^+(m)-\half}\left\{
\sum_{\ell=1}^{\wt\nu^{(j-1)}_{r}}
\wh f^{(r,m)}_{-1}(p)\,\exp(ip\lambda_{r,\ell}^{(j-1)})
+\sum_{\ell=1}^{\wt\nu^{(j)}_{r}}
\wh f_{2}^{(r,m)}(p)\,\exp(ip\lambda_{r,\ell}^{(j)})
+\sum_{\ell=1}^{\wt\nu^{(j+1)}_{r}}
\wh f_{-1}^{(r,m)}(p)\,\exp(ip\lambda_{r,\ell}^{(j+1)})
\right\}\nonumber
\end{eqnarray}
where
\begin{eqnarray}
\wh{f}_{-1}^{(r,m)}(p) &=& \frac{\sinh(\vert p\vert(\gamma^+(m)-\max(r,m)))
\sinh(\vert p\vert(\gamma^-(m)-\min(r,m)))}
{\sinh(\vert p\vert/2)\sinh(\vert p\vert(\gamma^+(m)-\gamma^-(m)))}\\
\wh{f}_{2}^{(r,m)}(p) &=& \frac{-2\sinh(\vert p\vert(\gamma^+(m)-\max(r,m)))
\sinh(\vert p\vert(\gamma^-(m)-\min(r,m)))}
{\tanh(\vert p\vert/2)\sinh(\vert p\vert(\gamma^+(m)-\gamma^-(m)))}-\delta_{r,m}
\end{eqnarray}
To obtain the explicit forms (\ref{eq:F1}) and (\ref{eq:F2}), one can perform the inverse 
Fourier transform by the theorem of residue and the integration. 
Remark that it is simpler to check these forms through a derivation and a direct 
Fourier transform. This fixes the equality
up to a constant. Considering the limit 
$\lambda_{0}\to\infty$, one shows that the constant vanishes.
\finprf

Finally, using relations (\ref{eq:correction}), (\ref{eq:bethebarN0}),  
(\ref{eq:bethebarN1}) and (\ref{eq:bethebarN2}) to simplify 
Bethe equations (\ref{eq:bethebarN}), we obtain the relation (\ref{eq:constraint}),
which ends the proof of theorem \ref{th:constraint}.


\begin{thebibliography}{99}

\bibitem{heisen}
W.~Heisenberg,
\textsl{Zur Theorie der Ferromagnetismus,}
Zeitschrift f{\"u}r Physik \textbf{49} (1928) 619.

\bibitem{bethe} 
H.~Bethe, 
\textsl{Zur Theorie der Metalle. Eigenwerte und Eigenfunktionen Atomkete,} 
Zeitschrift f{\"u}r Physik \textbf{71} (1931) 205.

\bibitem{comp1} M. Batchelor, X-W. Guan, N. Oelkers and A. Foerster,   
\textsl{Thermal and magnetic properties of integrable spin-1 and spin-3/2 
chains with applications to real compounds}, 
J. Stat. Mech. \textbf{0410} (2004) P017 and \texttt{cond-mat/0409311}.

\bibitem{condmat}  
S. Kimura, T. Takeuchi, K. Okunishi, M. Hagiwara, Z. He,  
K. Kindo, T. Taniyama and  M. Itoh,
\textsl{Novel ordering of an S = 1/2 quasi one-dimensional Ising-like 
anitiferromagnet in magnetic field}, \texttt{arXiv:0707.3713}.

\bibitem{carbon} 
Ling Ge, B. Montanari, J. Jefferson, D. Pettifor, N. Harrison,
G. Andrew and D. Briggs, 
\textsl{Modelling spin qubits in carbon peapods}, 
\texttt{arXiv:0710.3061}.

\bibitem{BS} 
N. Beisert and  M. Staudacher 
\textsl{Long-Range $PSU(2,2|4)$ Bethe Ansaetze for Gauge Theory 
and Strings}, Nucl.Phys. \textbf{B727} (2005) 1 and 
\texttt{hep-th/0504190}.

\bibitem{Agar} 
A. Agarwal 
\textsl{Aspects of Integrability in N =4 SYM},
Invited brief review for Mod. Phys. Lett. \textbf{A}, 
\texttt{arXiv:0708.2747}.

\bibitem{ZAFA}
A.~Zamolodchikov and V.~Fateev, 
\textsl{An integrable spin-1 Heisenberg chain,} Sov. J. Nucl. Phys. 
\textbf{32} (1980) 298.

\bibitem{Kul} P.~Kulish, N.~Reshetikhin and E.~Sklyanin,
\textsl{Yang-Baxter equation and representation theory: I},
Lett. Math. Phys. \textbf{5} (1981) 393.

\bibitem{Tak} L.~Takhtajan, 
\textsl{The picture of low-lying excitations in the isotropic
Heisenberg chain of arbitrary spins},
Phys. Lett. \textbf{A87} (1982) 479.

\bibitem{Bab} H.~Babujian, 
\textsl{Exact solution of the isotropic Heisenberg chain with arbitrary
spins: thermodynamics of the model}, Nucl. Phys. \textbf{B215} (1983)
317.

\bibitem{ow} E.~Ogievetsky and P.~Wiegmann,
\textsl{Factorized S-matrix and the Bethe Ansatz for simple Lie
groups,}
Phys. Lett. \textbf{168B} (1986) 360.

\bibitem{MENERI}
L.~Mezincescu, R.~Nepomechie and V.~Rittenberg,
\textsl{Bethe Ansatz solution of the Fateev-Zamolodchikov quantum
spin chain with boundary terms}, Phys. Lett. \textbf{A147} (1990) 70.

\bibitem{KoIzBo}
V.~Korepin, G.~Izergin and N.~Bogoliubov, 
\textsl{Quantum inverse
scattering method, correlation functions and algebraic Bethe Ansatz}
(Cambridge University Press, 1993).

\bibitem{fafa} L.~Faddeev, \textsl{How Algebraic Bethe Ansatz works
for integrable model}, Les Houches summerschool 1995 and
\texttt{hep-th/9605187}.

\bibitem{KuSu} A.~Kuniba and J.~Suzuki,
\textsl{Analytic Bethe Ansatz for fundamental representations of Yangians,}
Commun. Math. Phys. \textbf{173} (1995) 225.

\bibitem{Bytsko} A.~Bytsko, 
\textsl{On integrable Hamiltonians for higher spin XXZ chain,}
J. Math. Phys. \textbf{44} (2003) 3698 and
\texttt{hep-th/0112163}.

\bibitem{Tsuboi} Z.~Tsuboi ,
\textsl{From the quantum Jacobi--Trudi and Giambelli formula to a
nonlinear integral  equation for thermodynamics of the higher spin
Heisenberg model,}
J. Phys. \textbf{A37} (2004) 1747 and
\texttt{cond-mat/0308333}.

\bibitem{anjo}
N.~Andrei and H.~Johannesson,
\textsl{Heisenberg chain with impurities (an integrable model)},
Phys. Lett. \textbf{A100} (1984) 108.

\bibitem{fuka}
T.~Fukui and N.~Kawakami,
\textsl{Spin chains with periodic array of impurities},
Phys. Rev. \textbf{B55} (1997) R14709 and 
\texttt{cond-mat/9704072}.

\bibitem{YWang} Yupeng Wang,
\textsl{Exact solution of the open Heisenberg chain with two impurities,}
Phys. Rev. \textbf{B56} (1997) 14045
and \texttt{cond-mat/9805253}.

\bibitem{dewo}
H.~de~Vega and F.~Woynarovich, 
\textsl{New Integrable Quantum Chains combining different kind of spins}, 
J. Phys. \textbf{A25} (1992) 4499.

\bibitem{martins} S. Aladim and M.J. Martins, 
\textsl{Critical behaviour of integrable mixed-spin chains},
J. Phys. \textbf{A26} (1993) L529;\\
M.J. Martins, 
\textsl{The effects of a magnetic field in an integrable Heisenberg chain with mixed
spins}, J. Phys. \textbf{A26} (1993) 7301;\\
S.R. Aladim and M.J. Martins, 
\textsl{The class of universality of integrable and isotropic GL(N) mixed
magnets},
J. Phys. \textbf{A26} (1993) 7287 andÊÊ
\texttt{hep-th/9306049}.

\bibitem{abad2} 
J.~Abad and M.~Rios, 
\textsl{Integrable $su(3)$ spin chain combining different
representations}, \texttt{cond-mat/9706136}.

\bibitem{ana} A.~Doikou, 
\textsl{The XXX spin $s$ quantum chain and the alternating
$s^{1}$, $s^{2}$ chain with boundaries},
Nucl. Phys. \textbf{B634} (2002) 591 and \texttt{hep-th/0201008}.

\bibitem{KulResh} P.~Kulish and N.~Reshetikhin,   
\textsl{Diagonalisation of $GL(N)$ invariant transfer matrices and quantum 
N-wave system (Lee model)}, J. Phys. \textbf{A16} (1983) L591.

\bibitem{byebye} 
D.~Arnaudon, N.~Cramp\'e, A.~Doikou, L.~Frappat and {\'E}.~Ragoucy,
\textsl{Analytical Bethe Ansatz for closed and open $gl(n)$-spin chains
in any representation}, JSTAT \textbf{02} (2005) P02007,
\texttt{math-ph/0411021}.

\bibitem{MTV} E. Mukhin, V. Tarasov, and A. Varchenko,
\textsl{Bethe eigenvectors of higher transfer matrices}, 
J. Stat. Mech. \textbf{08} (2006) P08002
and \texttt{math.QA/0605015}.

\bibitem{TV} V. Tarasov and A. Varchenko,
\textsl{Combinatorial formulae for nested Bethe vectors},
\texttt{arXiv:math/0702277}.

\bibitem{paku} B. Enriquez, S. Khoroshkin and S. Pakuliak,
\textsl{Weight functions and Drinfeld currents}, Preprint ITEP-TH-40/05 
and \texttt{arXiv:math/0610398};\\
S. Khoroshkin,  S. Pakuliak and  V. Tarasov,
\textsl{Off-shell Bethe vectors and Drinfeld currents},
\texttt{arXiv:math/0610517}.

\bibitem{thermy} N. Cramp\'e, L. Frappat and \'E. Ragoucy,
\textsl{Thermodynamical limit of general gl(N) spin chains: vacuum state and densities,}
JSTAT \textbf{03} (2007) P03014 and \texttt{cond-mat/0701207}. 

\bibitem{sutherland} B.~Sutherland, 
\textsl{Model for a multicomponent quantum system,} 
Phys. Rev. \textbf{B12} (1975) 3795.

\bibitem{anne}
A. Doikou and R. Nepomechie,
\textsl{Bulk and boundary S matrices for the $su(N)$ chain,}
Nucl.Phys. \textbf{B521} (1998) 547 and \texttt{hep-th/980311}.

\bibitem{dVMN} H. de Vega,  L. Mezincescu and  R. Nepomechie,
\textsl{Scalar Kinks}, Int. J. Mod. Phys. \textbf{B8} (1994) 3473 and 
\texttt{hep-th/9402053}.

\bibitem{KS} P.~Kulish and E.~Sklyanin, 
\textsl{Quantum spectral transform method. Recent developments}, 
Lect. Notes in Phys. \textbf{151} (1982) 61.

\bibitem{TTF} V. Tarasov, L. Takhtadzhyan and L. Faddeev,
\textsl{Local Hamiltonians for integrable quantum models on a 
lattice}, Theor. Math. Phys. \textbf{57} (1983) 1059.

\end{thebibliography}
\end{document}